\documentclass[12pt,thmsa,a4paper,notitlepage]{article}
\usepackage{amsfonts}

\input{tcilatex}

\begin{document}

\author{M. H. LAGRAA$^{1,2}$\thanks{%
e-mail : meriem.lagraa@gmail.com }, M. LAGRAA$^{2}$\thanks{%
e-mail : m.lagraa@lycos.com} and N. TOUHAMI$^{2}\thanks{%
e-mail : touhami.nabila@gmail.com}$ \\
$^{1}$\textit{Ecole Pr\'{e}paratoire en Sciences et Techniques d'Oran
(EPSTO),}\\
\textit{\ B.P. 64 CH2 Achaba Hanifi, Oran, ALGERIA. }\\
$^{2}$\textit{Laboratoire de physique th\'{e}orique d'Oran (LPTO), }\\
\textit{Universit\'{e} d'Oran I, Ahmed Benbella, } \textit{B.P. 1524, El
M'Naouer,}\\
\textit{\ 31000 Es-S\'{e}nia, Oran, ALGERIA.}}
\title{On The Hamiltonian Formalism Of The Tetrad-Connection Gravity}
\maketitle

\begin{abstract}
We present a detailed analysis of the Hamiltonian constraints of the
d-dimensional tetrad-connection gravity where the non-dynamic part of the
spatial connection is fixed to zero by an adequate gauge transformation.
This new action leads to a coherent Hamiltonian formalism where the Lorentz,
scalar and vectorial first-class constraints obeying a closed algebra in
terms of Poisson brackets. This algebra closes with structure constants
instead of structure functions resulting from the Hamiltonian formalisms
based on the A.D.M. decomposition. The same algebra of the reduced
first-class constraints, where the second-class constraints are eliminated
as strong equalities, is obtained in terms of Dirac brackets. These
first-class constraints lead to the same physical degrees of freedom of the
general relativity.

\end{abstract}

\section{Introduction}

\bigskip

Any coherent canonical quantification of a theory requires a correct
treatment of its classical Hamiltonian formalism. During the last
half-century, canonical quantization of general relativity has attracted
much attention, especially these last few decades with the development of
the loop quantum gravity \cite{Ashtekar},\cite{Rovelli} and \cite{Thiemann}
(and references therein). Despite a lot of progress made in the different
approaches of the canonical quantization of the gravity, these approaches
are not complete in the sense that the algebra of the first-class
constraints closes with structure functions both in the metrical \cite%
{Arnowitt} and the tetrad formulation of gravity \cite{Ashtekar}. The
presence of structure functions can be a potential source of anomalies and
reveals that the first-class constraints do not correspond to symmetries
based on true Lie groups. This shows the special attention that must be paid
to the construction of the Hamiltonian formalism of gravity.

Among these different approaches, we cite the one where the Hamiltonian
formalism is derived from a generalized action \cite{holst} where the time
gauge is fixed in order to reduce Lorentz's manifest invariance of the
action to $SO(3)=SU(2)/Z_{2}$. The interest in $SO(3)$ came from the fact
that $SO(3)$ is compact allowing the construction of the Hilbert space in
loop quantum gravity. The Holst action does not modify the classical
Einstein equation in the vacuum but contains a new dimensionless parameter
known as the Barbero-Immirzi parameter \cite{Barb} which appears in the
spectra of area and volume operators at the quantum level \cite{Rovelli1}.
It also appears in the black hole entropy formula \cite{Rovelli2}. Note that
even at the classical level, once gravity is coupled to fermionic matter,
the Barbero-Immirzi parameter appears on-shell via the non zero torsion \cite%
{Perez}.

The local Lorentz invariance of the canonical vierbein form of general
relativity has been done in the second order formalism in \cite{Deser}. The
main complication of the covariant canonical formulation of the first order
tetrad-connection gravity is the presence of the second-class constraints
which require the Dirac brackets. The covariant Hamiltonian treatments of
the generalized $4-$dimensional action \cite{holst} have been developed
either in terms of Dirac brackets in \cite{Alexandrov}, where two copies of $%
su(2)$ Barbero tedrad-connection gravity are combined to get a $SO(4,C)$
covariant Hamiltonian, or in \cite{Barros} where the resolution of
second-class constraints leads to a reduced symplectic form where the phase
space elements obey canonical Poisson's brackets.

Almost at the same period, the enthusiasm aroused by the spin foam model of
the BF-theory \cite{Baez} and \cite{Barrett} has encouraged the
investigation of the covariant Hamiltonian of the tetrad gravity formulated
as a BF-theory with extra constraints on the $2-$form B. In \cite{Mariano}
the extra constraints on the $2-$form B are solved leading to similar
results as the ones of \cite{Barros}.

For higher dimensions, the analysis of the covariant Hamiltonian formalism
of the tetrad-connection action was performed in \cite{Bodendorfer} by
considering an extension of the A.D.M. phase space where the Lagrange
multipliers, the lapse and the shift, are considered as part of the phase
space. After solving the second-class constraints, a canonical reduced
symplectic form is obtained leading to an algebra of constraints which
closes with structure functions.

Until now all the Hamiltonian formulations of gravity where one starts from
the very beginning by the A.D.M. Decomposition of the tetrad components in
terms of lapse and shift lead to an algebra of first-class constraints
involving structure functions.

Rather than proceeding as in \cite{Bodendorfer}, we begin from the phase
space resulting directly from the tetrad-connection action without using the
A.D.M. formalism. We show that the connection splits in dynamic and non
dynamic parts. By fixing the non dynamic part of the connection to zero we
obtain a coherent Hamiltonian formalism where all the stages of the Dirac
procedure \cite{Dirac} for constrained systems are scrupulously respected.
Especially the second-class constraints are eliminated as strong equalities
only after the Dirac brackets are established.

The paper is organized as follows:

In section II, we investigate and classify the different constraints
according to Dirac terminology. The resolution of the equations involving
the Lagrange multiplier leads to a problematic constraint which is difficult
to check its consistency. In section III, we show that this problematic
constraint is tied to the non-dynamic part of the spatial connection which
can be fixed to zero by an adequate gauge fixing. The formalism derived from
the new action where the non-dynamic part of the connection is fixed to zero
leads to a consistent Hamiltonian treatment where the Lorentz, scalar and
vectorial first-class constraints form a Poisson algebra that closes with
structure constants. In section IV, we establish the Dirac brackets of the
reduced phase space elements where the second-class constraints are
eliminated as strong equalities. The reduced first-class constraints becomes
polynomial and obey, in terms of Dirac brackets, to the same algebra as the
one of the previous section and lead to the same physical degrees of freedom
of the general relativity. We end this section by showing that the solutions
of Hamiltonian's equations of the reduced phase space elements lead to a
zero torsion which is required to establish the equivalence between the
tetrad-connection gravity and the general relativity.

\section{Hamiltonian formalism of Tetrad-Gravity}

\bigskip In $d-$dimension space-time manifold $\mathcal{M}$, the functional
action of the tetrad-connection gravity is:

\begin{equation}
S(e,\omega )=\dint_{\mathcal{M}}\frac{1}{(d-2)!}e_{I_{0}}\wedge
...e_{I_{d-3}}\wedge \Omega _{I_{d-2}I_{d-1}}\epsilon ^{I_{0}...I_{d-1}}
\label{Eins-Cartan-Act}
\end{equation}%
\qquad where the capital Latin letters $I_{0},...,I_{d-1}\in \left[ 0,..,d-1%
\right] $ denote internal indices of the tensor representation spaces of the
Lorentz group, $\epsilon _{I_{0...}I_{d-1}}$ are the components of the
totally antisymmetric Levi-Cevita symbol, $\epsilon _{0...d-1}=-\epsilon
^{0...d-1}=1$, satisfying $\epsilon
_{I_{0}...I_{n}I_{n+1}...I_{d-1}}\epsilon
^{J_{0}...J_{n}I_{n+1}...I_{d-1}}=-(d-n)!\delta _{I_{0}...In}^{\left[
J_{0}...J_{n}\right] }$. $e_{I}=e_{\mu I}dx^{\mu }$ is the co-tetrad
one-form valued in the vectorial representation space endowed with the flat
metric $\eta _{IJ}=diag(-1,1,...,1)$ and $x^{\mu }$ are local coordinates of
the manifold $\mathcal{M}$ where the Greek letters $\mu ,$ $\nu $ $\in \left[
0,1,..,d-1\right] $ denote space-time indices ($t$ represents the time, $%
t=x^{0}=x^{t}$). Time-like indices will be labelled $"t"$ in the tangent
space and space-like indices will be labelled with small Latin letters $%
a,b,c\in \left[ 1,..,d-1\right] $. The metric $\eta _{IJ}$ and its inverse $%
\eta ^{IJ}$ are used to lower and to lift the Lorentz indices and to
determine the metric $g_{\mu \nu }=e_{\mu }^{I}e_{\nu }^{J}\eta _{IJ}$ of
the tangent space of the manifold $\mathcal{M}$. $\wedge $ is the wedge
product and $\Omega _{IJ}=-\Omega _{JI}=d\omega _{IJ}+\omega _{I}^{\text{ \ }%
N}\wedge \omega _{NJ}$ is the curvature two-form associated to the
connection one-form $\omega _{IJ}=-\omega _{JI}=\omega _{\mu IJ}dx^{\mu }$
valued in the $so(1,d-1)$ Lie-algebra. The co-tetrad and the connection are
supposed to be independent variables.

Before starting the Hamiltonian analysis of the action (\ref{Eins-Cartan-Act}%
), let us recall that in the Lagrangian formalism of fields theories the
basic variables are fields $\phi _{i}(x)$ and their time derivative $%
\partial _{t}\phi _{i}(x)$ which, in our case, are the co-tetrad components $%
e_{\mu I}$, the connection components $\omega _{\mu IJ}$ and their time
derivative $\partial _{t}\phi _{i}(x)=(\partial _{t}e_{\mu I},$ $\partial
_{t}\omega _{\mu IJ})$. These variables and their time derivatives are
considered as independent variables and constitute the configuration space.
In this framework the dynamic is presupposed determined by evolution
equations of second order with respect to time and the configuration space
is nothing but a space isomorph to the set of initial conditions of the
solutions of the evolution equations. It is not the case here where we are
dealing with a first order theory where the equations of motion are of the
first order, so they are only constraints in the configuration space. In
addition, since the action (\ref{Eins-Cartan-Act}) is given in terms of
differential forms, it follows that the time derivative of the temporal
components of the co-tetrad and the connection are absent from the evolution
equations that govern the tetrad-connection gravity theory and therefore the
evolution in time of these variables is undetermined. In the following,
these points will be investigated in the Hamiltonian formalism which is more
suitable for constrained systems.

To pass from the Lagrangian formalism to the Hamiltonian formalism, we will
suppose that the manifold $\mathcal{M}$ has topology $R\times \Sigma $,
where $R$ represents the time which is an evolution parameter of $d-1$
dimensional space-like hypersurfaces $\Sigma _{t}$ into the $d-$dimensional
manifold $\mathcal{M}$.\ In order to get the momenta conjugate to the
configuration fields $e_{\mu I}$ and $\omega _{\mu IJ}$ we must develop the
action (\ref{Eins-Cartan-Act}) in terms of components of the co-tetrad and
the connection

\begin{eqnarray}
S(e,\omega ) &=&\dint_{\mathcal{M}}\frac{1}{(d-2)!}e_{I_{0}}\wedge
...e_{I_{d-3}}\wedge \Omega _{I_{d-2}I_{d-1}}\epsilon ^{I_{0}...I_{d-1}} 
\nonumber \\
&=&\dint_{\mathcal{M}}\frac{1}{2(d-2)!}\epsilon ^{I_{0}...I_{d-1}}e_{\mu
_{0}I_{0}}...e_{\mu _{d-3}I_{d-3}}\Omega _{\mu _{d-2}\mu
_{d-1}I_{d-2}I_{d-1}}^{\text{ \ }}\epsilon ^{\mu _{0}...\mu _{d-1}}d^{d}x 
\nonumber \\
&=&-\dint_{\mathcal{M}}eA^{\mu K\nu L}\frac{\Omega _{\mu \nu KL}}{2}d^{d}x
\label{Einstein-Cartan1}
\end{eqnarray}%
where%
\[
\Omega _{IJ}=\frac{1}{2}\Omega _{\mu \nu IJ}dx^{\mu }\wedge dx^{\nu }=\frac{1%
}{2}\left( \partial _{\mu }\omega _{\nu IJ}-\partial _{\nu }\omega _{\mu
IJ}+\omega _{\mu I}^{\text{ \ \ \ }N}\omega _{\nu NJ}-\omega _{\nu I}^{\text{
\ \ \ }N}\omega _{\mu NJ}\right) dx^{\mu }\wedge dx^{\nu }, 
\]%
\begin{eqnarray}
eA^{\mu K\nu L} &=&\frac{1}{(d-2)!}\epsilon ^{I_{0}...I_{d-3}KL}e_{\mu
_{0}I_{0}}...e_{\mu _{d-3}I_{d-3}}\epsilon ^{\mu _{0}...\mu _{d-3}\mu \nu } 
\nonumber \\
&=&e(e^{\mu K}e^{\nu L}-e^{\nu K}e^{\mu L})=-A^{\nu K\mu L}=-A^{\mu L\nu K}%
\text{,}  \label{MatrixA}
\end{eqnarray}%
$e=\det (e_{\mu I})$, and $e^{\mu K}$ is the inverse of $e_{\mu L}$ , $%
e^{\mu K}e_{\mu L}=\delta _{L}^{K}$, $e^{\mu K}e_{\nu K}=\delta _{\nu }^{\mu
}$.

To carry out the Legendre transformations, the time derivatives must appear
explicitly in the action (\ref{Einstein-Cartan1})

\begin{equation}
S(e,\omega )=-\dint_{\mathcal{M}}\left( eA^{aKtL}\left( \partial _{t}\omega
_{aKL}-D_{a}\omega _{tKL}\right) -eA^{aKbL}\frac{\Omega _{abKL}}{2}\right)
d^{d}x  \label{Lagrangian}
\end{equation}%
from which we deduce the conjugate momenta $\pi ^{\beta N}$ and $\mathcal{P}%
^{\beta KL}$ of the co-tetrad $e_{\beta N}$ and the $so(1,d-1)$ connection $%
\omega _{\beta KL}$

\[
\pi ^{\beta N}(x)=\frac{\delta S(e,\omega )}{\delta \partial _{t}e_{\beta
N}(x)}=0\text{, }\mathcal{P}^{aKL}(x)=\frac{\delta S(e,\omega )}{\delta
\partial _{t}\omega _{aKL}(x)}=eA^{aKtL}(x) 
\]%
and 
\[
\mathcal{P}^{tKL}(x)=\frac{\delta S(e,\omega )}{\delta \partial _{t}\omega
_{tKL}(x)}=0 
\]%
obeying the following non-zero Poisson brackets

\begin{eqnarray}
\left\{ e_{\alpha I}(\overrightarrow{x}),\pi ^{\beta N}(\overrightarrow{y}%
)\right\} &=&\delta _{\alpha }^{\beta }\delta _{I}^{N}\delta (%
\overrightarrow{x}-\overrightarrow{y}),  \nonumber \\
\left\{ \omega _{\alpha IJ}(\overrightarrow{x}),\mathcal{P}^{\beta KL}\left( 
\overrightarrow{y}\right) )\right\} &=&\delta _{\alpha }^{\beta }\frac{1}{2}%
(\delta _{I}^{K}\delta _{J}^{L}-\delta _{I}^{L}\delta _{J}^{K})\delta (%
\overrightarrow{x}-\overrightarrow{y}).  \label{Poisson-Bracket}
\end{eqnarray}%
where $\overrightarrow{x}$ denotes the local coordinates $x^{a}$ of $\Sigma
_{t}$.

The expressions of the conjugate momenta lead to the following primary
constraints

\begin{eqnarray}
\pi ^{tN} &=&0,\text{ }\mathcal{P}^{tKL}=0\text{ },  \nonumber \\
\pi ^{bN} &=&0\text{ and }C^{aKL}=\mathcal{P}^{aKL}-eA^{aKtL}=0
\label{PrimaryConst}
\end{eqnarray}%
which satisfy the following non-zero Poisson brackets%
\begin{equation}
\left\{ \pi ^{aN}(\overrightarrow{x}),C^{bKL}(\overrightarrow{y})\right\}
=-eB^{aNtKbL}\delta (\overrightarrow{x}-\overrightarrow{y})
\label{Poisson-bracPI-C}
\end{equation}%
where $B^{\beta N\mu K\nu L}$ is defined as 
\begin{eqnarray}
eB^{\beta N\mu K\nu L} &=&\frac{1}{(d-3)!}\epsilon
^{I_{0}...I_{d-4}NKL}e_{\mu _{0}I_{0}}...e_{\mu _{d-4}I_{d-4}}\epsilon ^{\mu
_{0}...\mu _{d-4}\beta \mu \nu }  \nonumber \\
&=&e(e^{\beta N}A^{\mu K\nu L}+e^{\beta K}A^{\mu L\nu N}+e^{\beta L}A^{\mu
N\nu K})  \nonumber \\
&=&e(e^{\beta N}A^{\mu K\nu L}+e^{\mu N}A^{\nu K\beta L}+e^{\nu N}A^{\beta
K\mu L})=\frac{\delta }{\delta e_{\beta N}}eA^{\mu K\nu L}.  \label{MatrixB}
\end{eqnarray}%
\ \ 

The total Hamiltonian is defined by

\begin{equation}
\mathcal{H}_{T}=\int_{\Sigma }(\pi ^{tN}\Lambda _{tN}+\mathcal{P}^{tKL}\frac{%
\mathcal{A}_{tKL}}{2}+\pi ^{bN}\Lambda _{bN}+C^{aKL}\frac{\mathcal{A}_{aKL}}{%
2})+H_{0}  \label{TOTALHAM}
\end{equation}%
where $\Lambda _{tN}$, $\mathcal{A}_{tKL}$, $\Lambda _{bN}$ and $\mathcal{A}%
_{bKL}$ are the Lagrange multipliers for primary constraints (\ref%
{PrimaryConst}) and%
\[
H_{0}\mathcal{=}\int_{\Sigma }(eA^{aKbL}\frac{\Omega _{abKL}}{2}%
+eA^{aKtL}D_{a}\omega _{tKL}). 
\]

The consistency of the Hamiltonian formalism requires that these constraints
must be preserved under the time evolution given in term of total
Hamiltonian (\ref{TOTALHAM}) in the standard form:

\begin{equation}
\left\{ \pi ^{tN},\mathcal{H}_{T}\right\} =-eB^{tNaKbL}\frac{\Omega _{abKL}}{%
2}=P^{N}=0,  \label{PN}
\end{equation}

\begin{equation}
\left\{ \mathcal{P}^{tNM},\mathcal{H}_{T}\right\}
=D_{a}eA^{aNtM}=eB^{cKaNtM}D_{a}e_{cK}=M^{NM}=0,  \label{MKL}
\end{equation}

\begin{equation}
\left\{ \pi ^{bN},\mathcal{H}_{T}\right\} =-eB^{bNtKaL}\left( \frac{\mathcal{%
A}_{aKL}}{2}-D_{a}\omega _{tKL}\right) -eB^{bNaKcL}\frac{\Omega _{acKL}}{2}%
=0,  \label{EvolC1}
\end{equation}%
and%
\begin{eqnarray}
\left\{ C^{aKL},\mathcal{H}_{T}\right\} &=&eB^{bNtKaL}\left( \Lambda
_{bN}+\omega _{tN}^{\text{ \ \ }M}e_{bM}\right) +D_{c}(eA^{cKaL})  \nonumber
\\
&=&eB^{bNtKaL}\left( \Lambda _{bN}+\omega _{tN}^{\text{ \ \ }%
M}e_{bM}-D_{b}e_{tN}\right) +eB^{bNcKaL}D_{c}e_{bN}=0  \label{EvolC2}
\end{eqnarray}%
where we have used (\ref{Poisson-bracPI-C}) and $\left\{ \pi ^{\alpha
N},eA^{\mu K\nu L}\right\} =-\frac{\delta }{\delta e_{\alpha N}}eA^{\mu K\nu
L}=-eB^{\alpha N\mu K\nu L}$.

These consistency conditions show that the evolution of the constraint $\pi
^{tN}$ and $\mathcal{P}^{tNM}$ leads to the secondary constraints $P^{N}$
and $M^{NM}$ respectively, while the evolution of the constraints $\pi ^{bN}$
and $C^{aKL}$ leads to the equations for the Lagrange multipliers $\mathcal{A%
}_{aKL}\ $and $\Lambda _{aN}$.

Now we have to check the consistency of the secondary constraints. For the
constraint (\ref{MKL}) we get 
\begin{equation}
\left\{ M^{KL},\mathcal{H}_{T}\right\} =D_{a}(eB^{bNaKtL}\Lambda _{bN})+%
\frac{1}{2}\mathcal{A}_{aN}^{K}eA^{aNtL}+\frac{1}{2}\mathcal{A}%
_{aN}^{L}eA^{aKtN}.  \label{EvolLorenz}
\end{equation}

Using (\ref{EvolC2}) we obtain%
\begin{eqnarray*}
D_{a}\left( eB^{bNaKtL}\Lambda _{bN}\right) &=&-D_{a}\left(
eB^{bNtKaL}\Lambda _{bN}\right) \\
&=&D_{a}\left( eB^{bNtKaL}\omega _{tNM}e_{b}^{M}\right) +D_{a}\left( eB^{\mu
NbKaL}D_{b}e_{\mu N}\right)
\end{eqnarray*}%
where the second term of the right hand side of the equality above is
written as

\begin{eqnarray*}
D_{a}\left( eB^{\mu NbKaL}D_{b}e_{\mu N}\right) &=&D_{a}D_{b}eA^{bKal}=-%
\frac{1}{2}\left( D_{a}D_{b}-D_{b}D_{a}\right) eA^{aKbL} \\
&=&-\frac{1}{2}\left( \Omega _{ab\text{ \ }N}^{\text{ \ \ \ }%
K}eA^{aNbL}+\Omega _{ab\text{ \ }N}^{\text{ \ \ \ }L}eA^{aKbN}\right) \text{.%
}
\end{eqnarray*}

From the properties (\ref{MatrixB}) of the B-matrix, an explicit computation
leads, for any antisymmetric tensor $\mathcal{D}_{NM}=-\mathcal{D}_{MN},$ to
the identity%
\begin{eqnarray}
\mathcal{D}_{\text{ }N}^{K}eA^{\nu N\rho L}+\mathcal{D}_{\text{ }%
N}^{L}eA^{\nu K\rho N} &=&\frac{1}{2}(e_{\beta }^{K}eB^{\beta L\nu N\rho
M}-e_{\beta }^{L}eB^{\beta K\nu N\rho M})\mathcal{D}_{NM}  \nonumber \\
&=&\frac{1}{2}(e_{\beta }^{M}eB^{\beta N\nu K\rho L}-e_{\beta }^{N}eB^{\beta
M\nu K\rho L})\mathcal{D}_{NM}  \label{DKNANtL}
\end{eqnarray}%
leading to 
\begin{eqnarray*}
D_{a}\left( eB^{bNtKaL}\omega _{tNM}e_{b}^{M}\right) &=&D_{a}(\omega _{t%
\text{ }N}^{\text{ }K}eA^{tNaL}+\omega _{t\text{ }N}^{\text{ }L}eA^{tKaN}) \\
&=&-eA^{aNtL}D_{a}\omega _{t\text{ }N}^{\text{ }K}-eA^{aKtN}D_{a}\omega _{t%
\text{ }N}^{\text{ }L} \\
&&-\left( \omega _{t\text{ }N}^{\text{ }K}M^{NL}+\omega _{t\text{ }N}^{\text{
}L}M^{KN}\right)
\end{eqnarray*}%
from which we get 
\begin{eqnarray}
\left\{ M^{KL},\mathcal{H}_{T}\right\} &=&-\left( \omega _{t\text{ }N}^{%
\text{ }K}M^{NL}+\omega _{t\text{ }N}^{\text{ }L}M^{KN}\right)  \nonumber \\
&&-(\frac{\mathcal{A}_{aN}^{K}}{2}-D_{a}\omega _{t\text{ }N}^{\text{ }%
K})eA^{tNaL}-(\frac{\mathcal{A}_{aN}^{L}}{2}-D_{a}\omega _{t\text{ }N}^{%
\text{ }L})eA^{tKaN}  \nonumber \\
&&-\frac{1}{2}\left( \Omega _{ab\text{ \ }N}^{\text{ \ \ \ }%
K}eA^{aNbL}+\Omega _{ab\text{ \ }N}^{\text{ \ \ \ }L}eA^{aKbN}\right) .
\label{EvolLorenz1}
\end{eqnarray}

As a consequence of (\ref{DKNANtL}), (\ref{EvolLorenz1}) is written in the
form

\begin{eqnarray*}
&&\left\{ M^{KL},\mathcal{H}_{T}\right\} \\
&=&-\frac{1}{2}\left( e_{b}^{K}\left( eB^{bLtNaM}\left( \frac{1}{2}\mathcal{A%
}_{aNM}-D_{a}\omega _{tNM}\right) +\frac{1}{2}eB^{bLaNcM}\Omega
_{acNM}\right) \right) \\
&&+\left( K\leftrightarrow L\right) \\
&&+\frac{1}{2}\left( e_{t}^{K}P^{L}-e_{t}^{L}P^{K}\right) -\left( \omega _{t%
\text{ }N}^{\text{ }K}M^{NL}+\omega _{t\text{ }N}^{\text{ }L}M^{KN}\right)
\end{eqnarray*}%
which, when (\ref{EvolC1}) and (\ref{EvolC2}) are satisfied, reduces to%
\begin{equation}
\left\{ M^{KL},\mathcal{H}_{T}\right\} =\frac{1}{2}\left(
e_{t}^{K}P^{L}-e_{t}^{L}P^{K}\right) -\left( \omega _{t\text{ }N}^{\text{ }%
K}M^{NL}+\omega _{t\text{ }N}^{\text{ }L}M^{KN}\right) \simeq 0
\label{evollorentz1}
\end{equation}%
ensuring the consistency of the constraint $M^{KL}$. Here "$\simeq $"
denotes weak equality which means equality modulo the constraints.

The evolution of the constraint $P^{N}$ is given by%
\begin{equation}
\left\{ P^{N},\mathcal{H}_{T}\right\} =-eC^{aMtNbKcL}\frac{\Omega _{bcKL}}{2}%
\Lambda _{aM}-eB^{tNbKal}D_{b}\frac{\mathcal{A}_{aKL}}{2}  \label{evoltrans}
\end{equation}%
where

\begin{eqnarray}
&&eC^{\mu M\nu N\alpha K\beta L}=\frac{1}{(d-4)!}\epsilon
^{I_{0}...I_{d-5}MNKL}e_{\mu _{0}I_{0}}...e_{\mu _{d-5}I_{d-5}}\epsilon
^{\mu _{0}...\mu _{d-5}\mu \nu \alpha \beta }  \nonumber \\
&=&e^{\mu M}eB^{\nu N\alpha K\beta L}-e^{\beta M}eB^{\mu N\nu K\alpha
L}+e^{\alpha M}eB^{\beta N\mu K\nu L}-e^{\nu M}eB^{\alpha N\beta K\mu L} 
\nonumber \\
&=&e^{\mu M}eB^{\nu N\alpha K\beta L}-e^{\mu L}eB^{\nu M\alpha N\beta
K}+e^{\mu K}eB^{\nu L\alpha M\beta N}-e^{\mu N}eB^{\nu K\alpha L\beta M} 
\nonumber \\
&=&\frac{\delta eB^{\nu N\alpha K\beta L}}{\delta e_{\mu M}}.
\label{Cmatrix}
\end{eqnarray}

By using (\ref{EvolC1}), (\ref{EvolC2}) and the properties of the C-matrix (%
\ref{Cmatrix}) we get, after lengthy computation, the evolution of the
constraint $P^{N}$ under the form of combination of constraints 
\begin{eqnarray}
\left\{ P^{N},\mathcal{H}_{T}\right\} =-\omega _{t\text{ }M}^{\text{ }%
N}P^{M} &&  \nonumber \\
-e^{cN}\left( P^{M}\left( \Lambda _{cM}+\omega
_{tMK}e_{c}^{K}-D_{c}e_{tM}\right) +M^{KL}\left( \frac{\mathcal{A}_{cKL}}{2}%
-D_{c}\omega _{tKL}\right) \right) &\simeq &0.  \label{evolTrans1}
\end{eqnarray}

Note that the second term of the right hand side of (\ref{evolTrans1}) is
orthogonal to $e_{tN}$, therefore the evolution of $e_{tN}P^{N}$ gives%
\begin{equation}
\left\{ e_{tN}P^{N},\mathcal{H}_{T}\right\} =\left( \Lambda _{tN}+\omega
_{tN}^{\text{ \ \ }M}e_{tM}\right) P^{N}\simeq 0  \label{DetLamdaN}
\end{equation}%
while the part of (\ref{evolTrans1}) proportional to $e^{cN}$ shows, by
using the consistency of the constraint $M^{KL}$ (\ref{evollorentz1}),\ that
the evolution of the linear combination of the constraints $P^{N}$ and $%
M^{KL}$, $\mathcal{D}_{spa}=e_{aN}P^{N}+\omega _{aKL}M^{KL}$ is given by%
\begin{equation}
\left\{ e_{aN}P^{N}+\omega _{aKL}M^{KL},\mathcal{H}_{T}\right\}
=P^{N}\partial _{a}e_{tN}+M^{KL}\partial _{a}\omega _{tKL}\simeq 0.
\label{evoldiff}
\end{equation}

The consistency conditions (\ref{DetLamdaN}) and (\ref{evoldiff}) show that
contrary to the constraint $P^{N}$ the\ evolution of its projections $%
P^{N}e_{tN}$ and $\mathcal{D}_{spa}=e_{aN}P^{N}+\omega _{aKL}M^{KL}$ are
simple and independent of the Lagrange multipliers $\Lambda _{aN}$ and $%
\mathcal{A}_{aKL}$.

In what follows, instead of the constraint $P^{N}$ we consider its temporal
projection

\begin{equation}
\mathcal{D}_{t}=P^{N}e_{tN}=-eA^{aKbL}\frac{\Omega _{abKL}}{2}  \label{Dtemp}
\end{equation}%
and its smeared spatial projection%
\begin{equation}
\mathcal{D}_{sp}(\overrightarrow{N})=-\int_{\Sigma }N^{a}(e_{aN}P^{N}+\omega
_{aKL}M^{KL})=\int_{\Sigma }eA^{aKtL}\mathcal{L}_{\overrightarrow{N}}(\omega
_{aKL})  \label{Dspatial}
\end{equation}%
where we have used $eB^{tNaKbL}e_{tN}=eA^{aKbL}$ to obtain (\ref{Dtemp}) and 
\begin{eqnarray}
-N^{c}e_{cN}P^{N} &=&N^{c}e_{cN}eB^{tNaKbL}\frac{\Omega _{abKL}}{2}%
=eA^{aKtL}N^{c}\Omega _{caKL}  \nonumber \\
&=&eA^{aKtL}(\mathcal{L}_{\overrightarrow{N}}(\omega
_{aKL})-D_{a}(N^{c}\omega _{aKL}))  \label{EcNPN}
\end{eqnarray}%
\ to obtain (\ref{Dspatial}). $\mathcal{L}_{\overrightarrow{N}}(\omega
_{aKL})=N^{b}\partial _{b}\omega _{aKL}+\partial _{a}(N^{b})\omega _{bKL}$
is the Lie derivative along the arbitrary vector field $\overrightarrow{N}$
tangent to $\Sigma _{t}$. This Lie derivative does not affect the Lorentz
indices.

The relation (\ref{evoldiff}) exhibits the consistency of the constraint $%
\mathcal{D}_{sp}(\overrightarrow{N})$ as

\[
\left\{ \mathcal{D}_{sp}(\overrightarrow{N}),\mathcal{H}_{T}\right\}
=\int_{\Sigma }(P^{N}\mathcal{L}_{\overrightarrow{N}}(e_{tN})+M^{KL}\mathcal{%
L}_{\overrightarrow{N}}(\omega _{tKL}))\simeq 0 
\]%
which shows, by comparing to (\ref{evoldiff}), that $\mathcal{L}_{%
\overrightarrow{N}}(e_{tN})=N^{a}\partial _{a}e_{tN}$ and $\mathcal{L}_{%
\overrightarrow{N}}(\omega _{tKL})=N^{a}\partial _{a}\omega _{tKL}$ from
which we see that the Lie derivative $\mathcal{L}_{\overrightarrow{N}}$ \
treats the temporal components $e_{tN}$ and $\omega _{tKL}$ as scalars.

This analysis of the primary constraints $\pi ^{tN}$, $\mathcal{P}^{tKL}$, $%
\pi ^{aN}$ and $C^{aKL}$and the secondary constraints $\mathcal{D}_{t}$, $%
\mathcal{D}_{sp}$ and $M^{KL}$ shows that the set of constrains is complete
meaning that the total Hamiltonian $\mathcal{H}_{T}$ is coherent provided
that (\ref{EvolC1} ) and (\ref{EvolC2}) are satisfied.

To complete this analysis, we have to solve the equations (\ref{EvolC1}) and
(\ref{EvolC2}). We can check that 
\begin{eqnarray}
B_{\mu N\nu K\alpha L} &=&\frac{1}{2}\left( e_{\mu N}\frac{A_{\nu K\alpha L}%
}{d-2}+e_{\nu N}\frac{A_{\alpha K\mu L}}{d-2}+e_{\alpha N}A_{\mu K\nu
L}\right)  \nonumber \\
&=&\frac{1}{2}\left( \frac{A_{\mu N\nu K}}{d-2}e_{\alpha L}+\frac{A_{\mu
L\nu N}}{d-2}e_{\alpha K}+A_{\mu K\nu L}e_{\alpha N}\right)
\label{INVMATRIXB}
\end{eqnarray}%
is the inverse of $B^{\beta N\mu K\nu L}$ in the sense that 
\begin{equation}
B_{\mu N\nu K\alpha L}B^{\mu N\nu P\beta Q}=\delta _{\alpha }^{\beta
}(\delta _{K}^{P}\delta _{L}^{Q}-\delta _{L}^{P}\delta _{K}^{Q})
\label{inverse B1}
\end{equation}%
and%
\begin{equation}
B_{\mu N\nu K\alpha L}B^{\rho M\sigma K\alpha L}=\delta _{N}^{M}(\delta
_{\mu }^{\rho }\delta _{\nu }^{\sigma }-\delta _{\nu }^{\rho }\delta _{\mu
}^{\sigma }).  \label{inverseB2}
\end{equation}

We see from (\ref{INVMATRIXB}) that, contrary to $B^{\mu N\nu K\alpha L}$
which is antisymmetric with respect of the indices $\mu $, $\nu $ and $%
\alpha $ and of the indices $N$, $K$ and $L$, $B_{\mu N\nu K\alpha L}$ is
antisymmetric with respect of the indices $\mu $ and $\nu $ and of the
indices $K$and $L$ only.

For $\beta =b$ and $\alpha =a$, (\ref{inverse B1}) gives%
\begin{equation}
B_{cNtKaL}B^{cNtPbQ}+\frac{1}{2}B_{cNdKaL}B^{cNdPbQ}=\frac{1}{2}\delta
_{a}^{b}(\delta _{K}^{P}\delta _{L}^{Q}-\delta _{L}^{P}\delta _{K}^{Q})\text{%
.}  \label{AddProjecteur}
\end{equation}

As a consequence of the antisymmetric of the indices $\mu ,$ $\nu $ and $%
\alpha $ of $B^{\mu M\nu K\alpha L}$, (\ref{inverseB2}) gives for $\sigma
=\nu =t$ 
\begin{equation}
B_{cNtKaL}B^{bMtKaL}=\delta _{N}^{M}\delta _{c}^{b}.  \label{inverseB21}
\end{equation}%
and for $\sigma =t$ and $\nu =d$

\begin{equation}
B_{cNdKaL}B^{bMtKaL}=0.  \label{inverseB22}
\end{equation}

Using (\ref{inverseB21}) and (\ref{inverseB22}) we get the solution of (\ref%
{EvolC1}) as

\begin{equation}
\frac{1}{2}\mathcal{A}_{aKL}=D_{a}\omega _{tKL}-\frac{1}{2}%
B_{bNtKaL}B^{bNcPdQ}\Omega _{cdPQ}+B_{bNcKaL}\Lambda ^{bNc}
\label{MultAaKL1}
\end{equation}%
with arbitrary $\Lambda ^{bNc}$. The third term of the right hand side of (%
\ref{MultAaKL1}) is the part of solution of the homogeneous equation
associated with (\ref{EvolC1}).

The determination of the lagrange multipliers $\Lambda _{bN}$ is obtained by
multiplying (\ref{EvolC2}) by $B_{bNtKaL}$and using (\ref{inverseB21}) to get%
\begin{eqnarray}
\Lambda _{bN} &=&-\omega _{tN}^{\text{ \ \ }%
M}e_{bM}+D_{b}e_{tN}-B_{bNtKaL}B^{cMdK\alpha L}D_{d}e_{cM}  \nonumber \\
&=&-\omega _{tN}^{\text{ \ \ }M}e_{bM}+D_{b}e_{tN}+e^{-1}B_{bNtKtL}M^{KL}%
\simeq -\omega _{tN}^{\text{ \ \ }M}e_{bM}+D_{b}e_{tN}\text{.}
\label{MultiabN}
\end{eqnarray}

By multiplying (\ref{EvolC2}) by $B_{dNeKaL}$and using (\ref{inverseB22}) we
get%
\begin{eqnarray}
B_{dMaKeL}B^{bNcKeL}D_{c}e_{bN} &=&0=(\delta _{M}^{N}(\delta _{d}^{b}\delta
_{a}^{c}-\delta _{a}^{b}\delta _{d}^{c})-B_{dMaKtL}B^{bNcKtL})D_{c}e_{bN} 
\nonumber \\
&=&D_{a}e_{dM}-D_{d}e_{aM}-e^{-1}B_{dMaKtL}M^{KL}=0\text{.}
\label{TortionConstr}
\end{eqnarray}

The above condition is not a solution of the homogeneous equation associated
with (\ref{EvolC2}). It result neither from the Legendre transform nor from
the consistency of the constraints. This condition shows that the spatial
components of the torsion $\Theta _{Med}=\frac{1}{2}%
(D_{e}e_{dM}-D_{d}e_{eM}) $ are a combination of constraints $M^{KL}$. It is
a condition to have the general solution of (\ref{EvolC2}). In fact if we
multiply (\ref{MultiabN}) by $B^{bNtPeQ}$ and use (\ref{AddProjecteur}) we
get%
\begin{eqnarray*}
&&B^{bNtPeQ}\left( \Lambda _{bN}+\omega _{tN}^{\text{ \ \ }%
M}e_{bM}-D_{b}e_{tN}\right) +B^{bNtPeQ}\left(
B_{bNtKaL}B^{cMdKaL}D_{d}e_{cM}\right) \\
&=&B^{bNtPeQ}\left( \Lambda _{bN}+\omega _{tN}^{\text{ \ \ }%
M}e_{bM}-D_{b}e_{tN}\right) +B^{cMdPeQ}D_{d}e_{cM} \\
&&-\frac{1}{2}B^{bNfPeQ}B_{bNfKaL}B^{cMdKaL}D_{d}e_{cM}
\end{eqnarray*}%
implying, by virtue of (\ref{EvolC2}),%
\[
\frac{1}{2}B^{bNfPeQ}\left( B_{bNfKaL}B^{cMdKaL}D_{d}e_{cM}\right) =0 
\]%
which is equivalent to the condition (\ref{TortionConstr}).

By substituting (\ref{MultAaKL1}) and (\ref{MultiabN}) in $\mathcal{H}_{T}$,
we get 
\begin{eqnarray}
\mathcal{H}_{T}^{\prime } &=&\int_{\Sigma }\mathcal{P}^{tKL}\frac{\mathcal{A}%
_{tKL}}{2}+\pi ^{tN}\Lambda _{tN}  \nonumber \\
&&-\int_{\Sigma }((\pi ^{aK}e_{a}^{L}-\pi ^{aL}e_{a}^{K})+2D_{a}\mathcal{(}%
C^{aKL}+eA^{aKtL}))\frac{\omega _{tKL}}{2})  \nonumber \\
&&+\int_{\Sigma }(eA^{aKbL}\frac{\Omega _{abKL}}{2}+\pi
^{aN}D_{a}e_{tN}-C^{aKL}B_{bNtKaL}B^{bNcPdQ}\frac{\Omega _{cdPQ}}{2}) 
\nonumber \\
&&+\int_{\Sigma }C^{aKL}B_{bNcKaL}\Lambda ^{bNc}.  \label{hAMILTON1}
\end{eqnarray}

Now we check the consistency of constraints with the Hamiltonian $\mathcal{H}%
_{T}^{\prime }$. The evolution of the primary constraint $\pi ^{aN}$ is
consistent in the sense that its Poisson brackets with the Hamiltonian $%
\mathcal{H}_{T}^{\prime }$ give combinations of $\pi ^{aN}$ and $C^{aNM}$ as 
\begin{eqnarray}
\left\{ \pi ^{cN},\mathcal{H}_{T}^{\prime }\right\} &=&-\omega _{t\text{ }%
K}^{N}\pi ^{cK}  \nonumber \\
&&+\frac{1}{2}C^{aKL}\frac{\delta }{\delta e_{cN}}(B_{bMtKaL}B^{bMcPdQ})%
\frac{\Omega _{cdPQ}}{2}  \nonumber \\
&&-\frac{1}{2}C^{aKL}\frac{\delta }{\delta e_{cN}}(B_{bMdKaL})\Lambda
^{bMd}\simeq 0\text{.}  \label{prespiaN}
\end{eqnarray}

For the constraint $C^{aNM}$, we obtain

\begin{eqnarray}
\left\{ C^{aNM},\mathcal{H}_{T}^{\prime }\right\} &=&-\omega _{t\text{ }%
K}^{N}C^{aKM}-\omega _{t\text{ }K}^{M}C^{aNK}-\frac{1}{2}(\pi
^{aN}e_{t}^{M}-\pi ^{aM}e_{t}^{N})  \nonumber \\
&&+D_{b}(eA^{bNaM})+eB^{bQtNaM}D_{b}e_{tQ}-D_{c}(C^{dKL}B_{bNtKdL}B^{bNcNaM})
\label{presCaKL}
\end{eqnarray}%
which vanishes weakly if we use (\ref{TortionConstr}) to get%
\begin{eqnarray*}
D_{b}(eA^{bNaM})+eB^{bQtNaM}D_{b}e_{tQ} &=&\frac{1}{2}%
eB^{cQbNaM}(D_{b}e_{cQ}-D_{c}e_{bQ}) \\
&=&\frac{1}{2}eB^{cQbNaM}e^{-1}B_{cQbKtL}M^{KL}\simeq 0
\end{eqnarray*}%
showing that the evolution of the constraint $C^{aNM}$ is consistent only
when (\ref{TortionConstr}) is satisfied. This is due to the fact that (\ref%
{TortionConstr}) is a condition to solve the equation (\ref{EvolC2}) which
result from the consistency of the constraint $C^{aNM}$. From that, we
expect that the consistency of the constraints $M^{KL}$, $\mathcal{D}_{t}$
and $\mathcal{D}_{sp}(\overrightarrow{N})$ depends on (\ref{TortionConstr})
also. In fact a direct computation gives

\begin{eqnarray*}
\left\{ M^{KL},\mathcal{H}_{T}^{\prime }\right\} &=&-\omega
_{tN}^{K}M^{NL}-\omega _{tN}^{L}M^{KN}+\frac{1}{2}%
(e_{t}^{K}P^{L}-e_{t}^{L}P^{K}) \\
&&-\frac{1}{2}D_{a}(eB^{dNbKaL}(D_{b}e_{dN}-D_{d}e_{bN}))\text{,}
\end{eqnarray*}%
\[
\left\{ \mathcal{D}_{t},\mathcal{H}_{T}^{\prime }\right\} =-(\Lambda
_{N}+\omega _{tNM}e_{t}^{M})P^{N}+B_{dNcKbL}D_{a}(eA^{aKbL})\Lambda ^{dNc} 
\]%
and

\begin{eqnarray*}
\left\{ \mathcal{D}_{sp}(\overrightarrow{N}),\mathcal{H}_{T}^{\prime
}\right\} &=&-\int_{\Sigma }(P^{N}\mathcal{L}_{\overrightarrow{N}%
}e_{tN}+M^{KL}\mathcal{L}_{\overrightarrow{N}}\omega _{tKL}) \\
&&-\int_{\Sigma }\frac{1}{2}eB^{dNbKaL}(D_{b}e_{dN}-D_{d}e_{bN})\mathcal{L}_{%
\overrightarrow{N}}\omega _{aKL}
\end{eqnarray*}%
which show that the secondary constraints $M^{KL}$, $\mathcal{D}_{t}$, and $%
\mathcal{D}_{sp}(\overrightarrow{N})$ are consistent only when (\ref%
{TortionConstr}) is satisfied.

The consistency of the constraint $\pi ^{tN}$ gives 
\begin{eqnarray}
\left\{ \pi ^{tN},\mathcal{H}_{T}^{\prime }\right\} &=&-eB^{tNaKbL}\frac{%
\Omega _{abKL}}{2}+D_{a}\pi ^{aN}  \nonumber \\
&&+C^{aKL}(-e^{-1}e^{tN}B_{bMtKaL}eB^{bMdPeQ}  \nonumber \\
&&+e^{-1}\frac{\delta }{\delta e_{tN}}(B_{bMtKaL})eB^{bMdPeQ}  \nonumber \\
&&+e^{-1}B_{bMtKaL}eC^{tNbMdPeQ})\frac{\Omega _{dePQ}}{2}=P^{^{\prime
}N}\simeq 0  \label{PNprime}
\end{eqnarray}%
where we have used $\frac{\delta }{\delta e_{tN}}e^{-1}=-e^{-1}e^{tN}$ and (%
\ref{Cmatrix}).

From the relation $e_{tN}\frac{\delta }{\delta e_{tN}}%
(B_{bMtKaL})=B_{bMtKaL} $, the projection of the constraint $P^{\prime N}$, $%
\mathcal{D}_{t}^{\prime }=P^{\prime N}e_{tN}$, is written in a combination
of constraints 
\begin{equation}
\mathcal{D}_{t}^{\prime }=\mathcal{D}_{t}-\pi
^{aN}D_{a}e_{tN}+C^{aKL}B_{bNtKaL}B^{bNcPdQ}\frac{\Omega _{cdPQ}}{2}\simeq 0
\label{PprimN}
\end{equation}%
where we have used $eC^{tNbMdPeQ}e_{tN}=eB^{bMdPeQ}$ obtained from the
properties of the C-matrix (\ref{Cmatrix}).

For the projection on the spatial component of the co-tedrad we use the
relation $e_{cN}\frac{\delta }{\delta e_{tN}}(B_{bMtKaL})=B_{bMcKaL}$ and%
\[
e_{cN}C^{tNbMdPeQ}\frac{\Omega _{dePQ}}{2}=-B^{bMtPeQ}\Omega _{cePQ}+\delta
_{c}^{b}P^{M}\text{,} 
\]%
obtained from the properties of the C-matrix (\ref{Cmatrix}), to get%
\begin{eqnarray*}
&&C^{aKL}B_{bMtKaL}e_{cN}C^{tNbMdPeQ}\frac{\Omega _{dePQ}}{2} \\
&=&-C^{aKL}\Omega _{caKL}+\frac{1}{2}C^{aKL}B_{bMdKaL}B^{bMdPeQ}\Omega
_{cePQ}+C^{aKL}B_{cMtKaL}P^{M}
\end{eqnarray*}%
from which we obtain 
\begin{eqnarray*}
e_{cN}P^{\prime N} &=&-(C^{aKL}+eA^{aKtL})\Omega _{caKL}+e_{cN}D_{a}\pi
^{aN}+C^{aKL}e^{-1}B_{cMtKaL}P^{M} \\
&&+C^{aKL}(B_{bMcKaL}B^{bMdPeQ}\frac{\Omega _{dePQ}}{2}+B_{bMdKaL}B^{bMdPeQ}%
\frac{\Omega _{cePQ}}{2})
\end{eqnarray*}%
leading to a linear combination of smeared constraints as

\begin{eqnarray}
&&\dint_{\Sigma }N^{c}(e_{cN}P^{\prime N}+\frac{\omega _{cKL}}{2}M^{\prime
KL}+\pi ^{aN}(D_{a}e_{cN}-D_{c}e_{aN}))  \nonumber \\
&&-\dint_{\Sigma }N^{c}C^{aKL}(B_{bNcKaL}B^{bNdPeQ}\frac{\Omega _{dePQ}}{2}%
+B_{bNdKaL}B^{bNdPeQ}\frac{\Omega _{cePQ}}{2})  \nonumber \\
&&-\dint_{\Sigma }N^{c}C^{aKL}B_{cMtKaL}P^{M}  \nonumber \\
&=&-\dint_{\Sigma }(\pi ^{aM}\mathcal{L}_{\overrightarrow{N}%
}(e_{aM})+(C^{aKL}+eA^{aKtL})\mathcal{L}_{\overrightarrow{N}}(\omega
_{aKL})).  \label{DIFFSPPRIME}
\end{eqnarray}

Here the new secondary constraint $M^{\prime KL}$is deduced from the
consistency of the primary constraint $\mathcal{P}^{tNM}$

\begin{eqnarray}
\left\{ \mathcal{P}^{tKL},\mathcal{H}_{T}^{\prime }\right\} &=&(D_{a}%
\mathcal{(}C^{aKL}+eA^{aKtL})+\frac{1}{2}(\pi ^{aK}e_{a}^{L}-\pi
^{aL}e_{a}^{K}))  \nonumber \\
&=&\frac{1}{2}M^{\prime KL}\simeq 0\text{.}  \label{MKL1}
\end{eqnarray}

From the expressions (\ref{PprimN}) and (\ref{MKL1}), the total Hamiltonian
takes the compact form$\ $ 
\begin{eqnarray*}
\mathcal{H}_{T}^{\prime } &=&\int_{\Sigma }(\frac{1}{2}\mathcal{P}^{tKL}%
\mathcal{A}_{tKL}+\pi ^{tN}\Lambda _{tN}-\mathcal{D}_{t}^{\prime }-M^{\prime
KL}\frac{\omega _{tKL}}{2}) \\
&&+\dint_{\Sigma }C^{aKL}B_{bNcKaL}\Lambda ^{bNc}.
\end{eqnarray*}

The constraint (\ref{DIFFSPPRIME}) can be completed by adding the constraint 
$\pi ^{tN}\mathcal{L}_{\overrightarrow{N}}(e_{tN})$, where $\mathcal{L}_{%
\overrightarrow{N}}(e_{tN})=N^{a}\partial _{a}(e_{tN})$, to get 
\[
\mathcal{D}_{sp}^{\prime }(\overrightarrow{N})=\dint_{\Sigma }(\pi ^{\mu M}%
\mathcal{L}_{\overrightarrow{N}}(e_{\mu M})+(C^{aKL}+eA^{aKtL})\mathcal{L}_{%
\overrightarrow{N}}(\omega _{aKL})). 
\]%
which satisfy 
\[
\left\{ \mathcal{D}_{sp}^{\prime }(\overrightarrow{N}),\mathcal{D}%
_{sp}^{\prime }(\overrightarrow{N^{\prime }})\right\} =\mathcal{D}%
_{sp}^{\prime }(\left[ \overrightarrow{N},\overrightarrow{N^{\prime }}\right]
) 
\]%
where$\left[ \overrightarrow{N},\overrightarrow{N^{\prime }}\right] $ is the
Lie bracket. $\mathcal{D}_{sp}^{\prime }(\overrightarrow{N})$\ acts on the
co-tetrad components $e_{\mu N}$\ and on the connection $\omega _{aNM}$ as
diffeomorphisms of the hypersurface $\Sigma _{t}$ 
\begin{equation}
\left\{ e_{\mu N},\mathcal{D}_{sp}^{\prime }(\overrightarrow{N})\right\} =%
\mathcal{L}_{\overrightarrow{N}}(e_{\mu N})\text{, }\left\{ \omega _{aNM},%
\mathcal{D}_{sp}^{\prime }(\overrightarrow{N})\right\} =\mathcal{L}_{%
\overrightarrow{N}}(\omega _{aNM})\text{.}  \label{Constr-Tr-Tetr-Diff}
\end{equation}

The primary constraints $\pi ^{\mu N}$ and $C^{aNM}$ transform as%
\begin{equation}
\left\{ \pi ^{\mu N},\mathcal{D}_{sp}^{\prime }(\overrightarrow{N})\right\} =%
\mathcal{L}_{\overrightarrow{N}}(\pi ^{\mu N})\text{, }\left\{ C^{aNM},%
\mathcal{D}_{sp}^{\prime }(\overrightarrow{N})\right\} =\mathcal{L}_{%
\overrightarrow{N}}(C^{aNM})  \label{Constr-Trans-Diff}
\end{equation}%
which show that, contrary to the constraint $\mathcal{D}_{sp}(%
\overrightarrow{N})$, the Poisson brackets of $\mathcal{D}_{sp}^{\prime }(N)$
with the primary constraints $\pi ^{\mu N}$ and $C^{aNM}$\ vanish weakly. On
the other hand the above transformations imply that the constraints $%
\mathcal{D}_{t}^{\prime }$ and $M^{\prime KL}$are treated by the spatial
diffeomorphism constraint $\mathcal{D}_{sp}^{\prime }(\overrightarrow{N})$
as scalar densities of weight one%
\[
\left\{ \mathcal{D}_{sp}^{\prime }(\overrightarrow{N}),\mathcal{D}%
_{t}^{\prime }\right\} =-\mathcal{L}_{\overrightarrow{N}}(\mathcal{D}%
_{t}^{\prime })=-\partial _{a}(N^{a}\mathcal{D}_{t}^{\prime
})\Longrightarrow \left\{ \mathcal{D}_{sp}^{\prime }(\overrightarrow{N}),%
\mathcal{D}_{t}^{\prime }(M)\right\} =\mathcal{D}_{t}^{\prime }(\mathcal{L}_{%
\overrightarrow{N}}M) 
\]%
and%
\begin{eqnarray*}
\left\{ \mathcal{D}_{sp}^{\prime }(\overrightarrow{N}),M^{^{\prime }KL}(%
\overrightarrow{x})\right\} &=&-\mathcal{L}_{\overrightarrow{N}}(M^{^{\prime
}KL}(\overrightarrow{x}))=-\partial _{a}(N^{a}M^{^{\prime }KL}(%
\overrightarrow{x})) \\
&\Longrightarrow &\left\{ \mathcal{D}_{sp}^{\prime }(\overrightarrow{N}),%
\mathcal{M}^{\prime }(\theta )\right\} =\mathcal{M}^{\prime }(\mathcal{L}_{%
\overrightarrow{N}}\theta )
\end{eqnarray*}%
which can be verified by a direct computation. Here $\mathcal{D}_{t}^{\prime
}(M)=\dint_{\Sigma }M\mathcal{D}_{t}^{\prime }$ is the smeared scalar
constraint where $M$ is an arbitrary function and $\mathcal{M}^{\prime
}(\theta )=\dint_{\Sigma }M^{\prime KL}\frac{\theta _{KL}}{2}$ where $\theta
_{KL}$ may be identified to the dimensionless infinitesimal arbitrary
parameters $\theta _{KL}=\delta t\omega _{tKL}$.

From (\ref{Constr-Tr-Tetr-Diff}) and (\ref{Constr-Trans-Diff}), We \ deduce 
\[
\left\{ \mathcal{D}_{sp}^{\prime }(\overrightarrow{N}),\int_{\Sigma
}C^{aKL}B_{bNcKaL}\Lambda ^{bNc}\right\} =\int_{\Sigma }C^{aKL}B_{bNcKaL}%
\mathcal{L}_{\overrightarrow{N}}(\Lambda ^{bNc})\simeq 0 
\]%
showing that the constraint $\mathcal{D}_{sp}^{\prime }(\overrightarrow{N})$
is preserved in the time evolution. In addition, the Poisson bracket of $%
\mathcal{D}_{sp}^{\prime }(\overrightarrow{N})$ with the condition (\ref%
{TortionConstr}) gives%
\[
\left\{ \mathcal{D}_{sp}^{\prime }(\overrightarrow{N}%
),B_{dMaKeL}B^{bNcKeL}D_{c}e_{bN}\right\} =\mathcal{L}_{\overrightarrow{N}%
}(B_{dMaKeL}B^{bNcKeL}D_{c}e_{bN})\simeq 0 
\]%
\ which shows that $\mathcal{D}_{sp}^{\prime }(\overrightarrow{N})$ is a
first-class constraint.

We may also complete the Lorentz constraint $M^{\prime KL}$ by adding the
constraint $\pi ^{tN}$ as

\[
\frac{M^{\prime KL}}{2}=(D_{a}\mathcal{(}C^{aKL}+eA^{aKtL})+\frac{1}{2}(\pi
^{\mu K}e_{\mu }^{L}-\pi ^{\mu L}e_{\mu }^{K})) 
\]%
which acts on the co-tetrad components $e_{\mu N}$\ and\ on the connection $%
\omega _{aNM}$ like local infinitesimal transformations of gauge

\begin{equation}
\left\{ e_{\mu N},\mathcal{M}^{\prime }(\theta )\right\} =\theta _{N}^{\text{
\ \ }L}e_{\mu L}\text{, }\left\{ \omega _{aNM},\mathcal{M}^{\prime }(\theta
)\right\} =-D_{a}\theta _{NM}\text{.}  \label{PH-SP6LTR1}
\end{equation}

The primary constraints $\pi ^{\mu N}$ and $C^{aNM}$ transform like the
contravariant tensors%
\begin{equation}
\left\{ \pi ^{\mu N},\mathcal{M}^{\prime }(\theta )\right\} =\theta _{\text{
\ }L}^{N}\pi ^{\mu L}\text{, }\left\{ C^{aNM},\mathcal{M}^{\prime }(\theta
)\right\} =\theta _{\text{ \ }L}^{N}C^{aLM}+\theta _{\text{ \ }L}^{M}C^{aNL}
\label{PH-SPLTR2}
\end{equation}%
from which we deduce that the Poisson brackets of $\mathcal{M}^{\prime
}(\theta )$ with the primary constraints $\pi ^{aN}$ and $C^{aNM}$ vanish
weakly. The fact that the space-time indices do not transform facilitate the
calculation of transformations that $\mathcal{M}^{\prime }(\theta )$
generates.

The constraint $\mathcal{D}_{t}$ is a scalar under Lorentz transformations%
\[
\left\{ \mathcal{D}_{t}(\overrightarrow{x}),\mathcal{M}^{\prime }(\theta
)\right\} =0\Longrightarrow \left\{ \mathcal{D}_{t}(\overrightarrow{x}),%
\mathcal{M}^{\prime }(\theta )\right\} =0 
\]%
and $M^{\prime KL}$ is a contravariant tensor 
\[
\left\{ M^{\prime KL},\mathcal{M}^{\prime }(\theta )\right\} =\theta _{\text{
}N}^{K}M^{\prime NL}+\theta _{\text{ }N}^{L}M^{\prime KN} 
\]%
leading to the $so(1,d-1)$ Lie algebra%
\begin{eqnarray*}
\left\{ M^{\prime NM}(\overrightarrow{x}),M^{\prime KL}(\overrightarrow{y}%
)\right\} &=&(\eta ^{NL}M^{\prime MK}(\overrightarrow{x})+\eta
^{MK}M^{\prime NL}(\overrightarrow{x}) \\
&&-\eta ^{NK}M^{\prime ML}(\overrightarrow{x})-\eta ^{ML}M^{\prime NK}(%
\overrightarrow{x}))\delta (\overrightarrow{x}-\overrightarrow{y}).
\end{eqnarray*}

The transformations rules (\ref{PH-SP6LTR1}) and (\ref{PH-SPLTR2}) lead to 
\[
\left\{ \mathcal{M}^{\prime }(\theta ),\int_{\Sigma
}C^{aKL}B_{bNcKaL}\Lambda ^{bNc}\right\} =\int_{\Sigma }\theta _{N}^{\text{
\ }M}C^{aKL}B_{bMcKaL}\Lambda ^{bNc} 
\]%
and 
\[
\left\{ \mathcal{M}^{\prime }(\theta
),B_{dMaKeL}B^{bNcKeL}D_{c}e_{bN}\right\} =\theta _{M}^{\text{ \ \ \ }%
Q}B_{dQaKeL}B^{bNcKeL}D_{c}e_{bN} 
\]%
showing that $M^{\prime KL}$ is first-class constraint.

Finally, a straightforward computation gives

\[
\left\{ \mathcal{D}_{t}^{\prime }(M),\mathcal{D}_{t}^{\prime }(M^{\prime
})\right\} =0\text{.} 
\]

But (\ref{presCaKL}) shows that the Poisson brackets of $\mathcal{D}%
_{t}^{\prime }(M)$ with the primary constraints $C^{aNM}$ and with $%
\int_{\Sigma }C^{aKL}B_{bNcKaL}\Lambda ^{bNc}$ vanish modulo the constraint (%
\ref{TortionConstr}). Therefore, the scalar constraint $\mathcal{D}%
_{t}^{\prime }$ is preserved under the time evolution and can be considered
as a first-class constraint only if the condition (\ref{TortionConstr}) is
satisfied.

In conclusion, we are in presence of a Hamiltonian formalism of the
tetrad-connection gravity composed of first-class constraints, $\pi ^{tN},$ $%
\mathcal{P}^{tNM}$, the Lorentz constraint$\ \mathcal{M}^{\prime }(\theta )$
and the spatial diffeomorphism constraint $\mathcal{D}_{sp}^{\prime }(%
\overrightarrow{N})$. Although the scalar constraint $\mathcal{D}%
_{t}^{\prime }(M)$ forms with $\mathcal{M}^{\prime }(\theta )$\ and $%
\mathcal{D}_{sp}^{\prime }(\overrightarrow{N})$ a closed algebra with
structure constants, its Poisson bracket with the primary constraint $%
C^{aNM} $ vanishes weakly only if the condition (\ref{TortionConstr}) is
resolved.

\section{The fixing of the non-dynamical connection}

In spite of the fact that we have obtained a closed algebra in terms of
structure constants, the constraint (\ref{TortionConstr}) is problematic
because of the difficulties to check its consistency. To avoid this problem,
we decompose the spatial connection as

\[
\omega _{aKL}=\omega _{1aKL}+\omega _{2aKL} 
\]%
where%
\[
\omega _{1aKL}=P_{1KaL}^{\text{ \ \ \ \ \ \ \ }PdQ}\omega
_{dPQ}=B_{bNtKaL}B^{bNtPdQ}\omega _{dPQ} 
\]%
and%
\[
\omega _{2aKL}=P_{2KaL}^{\text{ \ \ \ \ \ \ \ }PdQ}\omega _{dPQ}=\frac{1}{2}%
B_{bNcKaL}B^{bNcPdQ}\omega _{dPQ}. 
\]

It is easy to check from (\ref{AddProjecteur}), (\ref{inverseB21}) and (\ref%
{inverseB22}) that%
\begin{eqnarray*}
P_{1KaL}^{\text{ \ \ \ \ \ \ \ }PdQ}+P_{2KaL}^{\text{ \ \ \ \ \ \ \ }PdQ} &=&%
\frac{1}{2}\delta _{a}^{d}\left( \delta _{K}^{P}\delta _{L}^{Q}-\delta
_{L}^{P}\delta _{K}^{Q}\right) \text{,} \\
P_{1KaL}^{\text{ \ \ \ \ \ \ \ }NbM}P_{1NbM}^{\text{ \ \ \ \ \ \ \ }PdQ}
&=&P_{1KaL}^{\text{ \ \ \ \ \ \ \ }PdQ}\text{, }P_{2KaL}^{\text{ \ \ \ \ \ \
\ }NbM}P_{2NbM}^{\text{ \ \ \ \ \ \ \ }PdQ}=P_{2Kal}^{\text{ \ \ \ \ \ \ \ }%
PdQ}
\end{eqnarray*}%
and

\[
P_{1KaL}^{\text{ \ \ \ \ \ \ \ }NbM}P_{2NbM}^{\text{ \ \ \ \ \ \ \ }PdQ}=0 
\]%
which show that $P_{1KaL}^{\text{ \ \ \ \ \ \ \ }PdQ}$ and $P_{2KaL}^{\text{
\ \ \ \ \ \ \ }PdQ}$ are projectors.

These projections of the connection is motivated by the fact that the time
derivative of $\omega _{2aKL}$ does not contribute to the kinematic part of
the action (\ref{Lagrangian}). In fact, from the identities%
\[
B_{bNcKaL}e^{aL}=B_{bNcKaL}e^{aK}=B_{bNcKaL}e^{tL}=B_{bNcKaL}e^{tK}=0 
\]%
we deduce the relations%
\[
eA^{aKtL}P_{2KaL}^{\text{ \ \ \ \ \ \ \ }PdQ}=0\text{ and }%
eA^{aKtL}P_{1KaL}^{\text{ \ \ \ \ \ \ \ }PdQ}=eA^{dPtQ} 
\]%
from which we get 
\begin{eqnarray*}
eA^{aKtL}\partial _{t}\omega _{2aKL} &=&\partial _{t}(eA^{aKtL}\omega
_{2aKL})-\partial _{t}(eA^{aKtL})\omega _{2aKL} \\
&=&-eB^{bNtKaL}(\partial _{t}e_{bN})\omega _{2aKL}=0
\end{eqnarray*}%
as a consequence of (\ref{inverseB22}). Therefore, like for the temporal
component of the co-tetrad and of the connection, the projected spatial
connection $\omega _{2aKL}$ is non-dynamic.

We also have

\[
eA^{aKtL}D_{2a}\omega _{tKL}=eA^{aKtL}P_{2KaL}^{\text{ \ \ \ \ \ \ }%
PdQ}D_{d}\omega _{tPQ}=0 
\]%
implying 
\[
eA^{aKtL}D_{a}\omega _{tKL}=eA^{aKtL}P_{1KaL}^{\text{ \ \ \ \ \ \ }%
PdQ}D_{d}\omega _{tPQ}=eA^{aKtL}D_{1a}^{\omega _{1}}\omega _{tKL} 
\]%
which shows that only the projected part $P_{1KaL}^{\text{ \ \ \ \ \ \ \ }%
PdQ}D_{d}\omega _{tPQ}=D_{1a}^{\omega _{1}}\omega _{tKL}$ of the covariant
derivative of $\omega _{tKL}$ given in terms of the connection $\omega
_{1aKL}$ contributes to the action. So, the action (\ref{Lagrangian}) can be
rewritten under the form

\begin{equation}
S(e,\omega )=\dint_{\mathcal{M}}\left( eA^{aKtL}\left( \partial _{t}\omega
_{1aKL}-D_{1a}^{\omega _{1}}\omega _{tKL}\right) -eA^{aKbL}\frac{\Omega
_{abKL}}{2}\right)  \label{Reduc-action0}
\end{equation}%
showing that the two parts of the spacial connection do not play the same
role. The non-dynamic spatial connection $\omega _{2aKL}$\ contributes only
to the third term. The above section showed us that the temporal components
of the connection are Lagrange multipliers and $\delta t\omega _{tKL}=\theta
_{KL}$ play the role of infinitesimal dimensionless parameters of local
transformations of the Lorentz group under which the spatial connection
transforms as $\delta \omega _{aKL}=$ $-D_{a}\theta _{KL}=D_{a}\delta
t\omega _{tKL}$. Since the projected part $D_{2a}\omega _{tKL}$ does not
contribute to the action, we can fix it to zero without modifying the action
(\ref{Lagrangian}). We will show that this gauge fixing results from the
fixing of the non-dynamic connection\ $\omega _{2aKL}=0$.

Before showing how the gauge transformations of the connection allow us to
fix the non-dynamic connection $\omega _{2aKL}$ to zero, let us note that
the ranks of the propagators $P_{1KaL}^{\text{ \ \ \ \ \ \ \ }PdQ}$ and $%
P_{2KaL}^{\text{ \ \ \ \ \ \ \ }PdQ}$, given by their trace, 
\[
\frac{1}{2}\delta _{d}^{a}\left( \delta _{P}^{K}\delta _{Q}^{L}-\delta
_{P}^{K}\delta _{P}^{L}\right) P_{1KaL}^{\text{ \ \ \ \ \ \ \ }PdQ}=d(d-1) 
\]%
and 
\[
\frac{1}{2}\delta _{d}^{a}\left( \delta _{P}^{K}\delta _{Q}^{L}-\delta
_{P}^{K}\delta _{P}^{L}\right) P_{2KaL}^{\text{ \ \ \ \ \ \ \ }PdQ}=\frac{1}{%
2}d(d-1)(d-3) 
\]%
are equal exactly to the number of independent components $\omega _{1aKL}$
and $\omega _{2aKL}$ respectively. On the other side the number $\frac{1}{2}%
d(d-1)(d-3)$ of independent relations (\ref{TortionConstr}) which is equal
to $\frac{1}{2}d(d-1)(d-2)$ relations in (\ref{TortionConstr}) minus $\frac{1%
}{2}d(d-1)$ identities%
\[
eB^{dNcPtQ}\left( D_{c}e_{dN}-D_{d}e_{cN}-e^{-1}B_{dNcKtL}M^{KL}\right)
=2M^{PQ}-2M^{PQ}=0 
\]%
corresponds exactly to the number of components $\omega _{2aKL}$. Since the
equation which results from the functional derivative of the action (\ref%
{Reduc-action0}) with respect to $\omega _{2aKL}$ is (\ref{TortionConstr}),
an action which does not contain explicitly this projected spacial
connection does not lead to (\ref{TortionConstr}).

Now, let us see how to fix the non-dynamic spatial connection. From the
Lorentz infinitesimal transformations $\delta e_{\mu K}=\theta _{K}^{\text{
\ }N}e_{\mu N}$ and $\delta \omega _{aKL}=-D_{a}\theta _{KL}$ we deduce

\[
\text{\ }\delta \omega _{1aKL}=\theta _{K}^{\text{ \ }N}\omega
_{1aNL}+\theta _{L}^{\text{ \ }N}\omega _{1aKN}-P_{1KaL}^{\text{ \ \ \ \ \ \
\ }PdQ}\partial _{d}\theta _{PQ} 
\]%
and 
\[
\delta \omega _{2aKL}=\theta _{K}^{\text{ \ }N}\omega _{2aNL}+\theta _{L}^{%
\text{ \ }N}\omega _{2aKN}-P_{2KaL}^{\text{ \ \ \ \ \ \ \ }PdQ}\partial
_{d}\theta _{PQ} 
\]%
which show that each part of the projected spatial connection transforms
independently of the other. This allows us to fix the non-dynamic part of
the connection to zero 
\[
\omega _{2aKl}^{\prime }=\omega _{2aKL}+\theta _{K}^{\text{ \ }N}\omega
_{2aNL}+\theta _{L}^{\text{ \ }N}\omega _{2aKN}-P_{2KaL}^{\text{ \ \ \ \ \ \
\ }PdQ}\partial _{d}\theta _{PQ}=0\text{.} 
\]

A transformation of $\omega _{2aKl}^{\prime }$\ gives%
\begin{eqnarray*}
\omega _{2aKl}^{^{\prime \prime }} &=&\omega _{2aKL}^{\prime }+\theta
_{K}^{^{\prime }\text{ \ }N}\omega _{2aNL}^{\prime }+\theta _{L}^{^{\prime }%
\text{ \ }N}\omega _{2aKN}^{\prime }-P_{2KaL}^{\text{ \ \ \ \ \ \ \ }%
PdQ}\partial _{d}\theta _{PQ}^{\prime } \\
&=&-P_{2KaL}^{\text{ \ \ \ \ \ \ \ }PdQ}\partial _{d}\theta _{PQ}^{\prime }
\end{eqnarray*}%
showing that this fixing of the non-dynamic part of the connection remains
invariant if%
\begin{equation}
\text{\ }P_{2KaL}^{\text{ \ \ \ \ \ \ \ }PdQ}\partial _{d}\theta
_{PQ}^{\prime }=\partial _{2a}\theta _{KL}^{\prime }=0\Longrightarrow
\partial _{a}\theta _{KL}=\text{\ }P_{1KaL}^{\text{ \ \ \ \ \ \ \ }%
PdQ}\partial _{d}\theta _{PQ}\text{.}  \label{CondLorentz2}
\end{equation}

On the other hand, in order to keep the same degrees of freedom during gauge
transformations of $\omega _{1aKL}$, we impose

\begin{equation}
\text{\ }\delta _{2}\omega _{1aKL}=P_{2KaL}^{\text{ \ \ \ \ \ \ \ }%
PdQ}\delta \omega _{1dPQ}=-P_{2KaL}^{\text{ \ \ \ \ \ \ \ }PdQ}(\theta _{P}^{%
\text{ \ }N}\omega _{1aNQ}+\theta _{Q}^{\text{ \ }N}\omega _{1aPN})=0
\label{CondLorentz1}
\end{equation}%
implying 
\[
\text{\ }(\theta _{K}^{\text{ \ }N}\omega _{1aNL}+\theta _{L}^{\text{ \ }%
N}\omega _{1aKN})=P_{1KaL}^{\text{ \ \ \ \ \ \ \ }PdQ}(\theta _{P}^{\text{ \ 
}N}\omega _{1aNQ}+\theta _{Q}^{\text{ \ }N}\omega _{1aPN})=0\text{.} 
\]

The relations (\ref{CondLorentz2}) and (\ref{CondLorentz1}) show that the
fixing of the non-dynamic connection to zero does not restrict the gauge
parameters but only the gauge transformations of the dynamic part of the
spatial connection to%
\begin{equation}
\delta e_{\mu K}=\theta _{K}^{\text{ \ }N}e_{\mu N}\text{ \ , }\delta \omega
_{tNM}=-D_{t}\theta _{NM}\text{ and \ }\delta \omega _{1aKL}=-D_{1a}^{\omega
_{1}}\theta _{KL}  \label{TransLorentzPROJ1}
\end{equation}%
subject to the condition%
\begin{equation}
\text{\ }D_{2a}^{\omega _{1}}\theta _{KL}=0\text{ and }\partial _{2a}\theta
_{KL}=0\text{.}  \label{gauge-restrict}
\end{equation}

Since the first and the second term of (\ref{Reduc-action0}) do not depend
of $\omega _{2aKL}$, to verify that the Lagrangian density (\ref%
{Reduc-action0}), where $\omega _{2aKL}=0$, is invariant under the
infinitesimal gauge transformations (\ref{TransLorentzPROJ1}) subject to the
conditions (\ref{gauge-restrict}), it suffices to check the invariance of
the third term 
\begin{eqnarray}
&&\delta \left( eA^{aKbL}\frac{\Omega (\omega _{1})_{abKL}}{2}\right)
=(\theta _{\text{ }N}^{K}eA^{aNbL}+\theta _{\text{ }N}^{L}eA^{aKbN})\frac{%
\Omega (\omega _{1})_{abKL}}{2}  \nonumber \\
&&-eA^{aKbL}D_{a}^{\omega _{1}}D_{1b}^{\omega _{1}}\theta _{KL}  \nonumber \\
&=&(\theta _{\text{ }N}^{K}eA^{aNbL}+\theta _{\text{ }N}^{L}eA^{aKbN})\frac{%
\Omega (\omega _{1})_{abKL}}{2}-eA^{aKbL}D_{a}^{\omega _{1}}D_{b}^{\omega
_{1}}\theta _{KL}  \nonumber \\
&=&(\theta _{\text{ }N}^{K}eA^{aNbL}+\theta _{\text{ }N}^{L}eA^{aKbN})\frac{%
\Omega (\omega _{1})_{abKL}}{2}  \nonumber \\
&&-eA^{aKbL}\left( \frac{\Omega (\omega _{1})_{abKN}}{2}\theta _{\text{ \ }%
L}^{N}+\frac{\Omega (\omega _{1})_{abLN}}{2}\theta _{K}^{\text{ \ }N}\right)
=0\text{.}  \label{Inv-H0fixed}
\end{eqnarray}

In what follows, we consider the action (\ref{Lagrangian}) where we fix $%
\omega _{2aKL}$ to zero:%
\begin{equation}
S^{f}(e,\omega _{1})=\dint_{\mathcal{M}}\left( eA^{aKtL}\left( \partial
_{t}\omega _{1aKL}-D_{1a}^{\omega _{1}}\omega _{tKL}\right) -eA^{aKbL}\frac{%
\Omega (\omega _{1})_{abKL}}{2}\right) \text{.}  \label{Reduc-action1}
\end{equation}

The fixed phase space, $e_{tN}$, $e_{aN}$, $\omega _{tKL}$, $\omega _{1aKL}$
and their conjugate momenta $\pi ^{tN}$, $\pi ^{aN}$, $\mathcal{P}^{tKL}$\
and $\mathcal{P}_{1}^{aKL}=P_{PbQ}^{\text{ \ \ \ \ \ }KaL}\mathcal{P}^{bPQ}$%
, is equipped by the following non zero Poisson brackets

\begin{eqnarray*}
\left\{ e_{\alpha I}(\overrightarrow{x}),\pi ^{\beta N}(\overrightarrow{y}%
)\right\} &=&\delta _{\alpha }^{\beta }\delta _{I}^{N}\delta (%
\overrightarrow{x}-\overrightarrow{y}), \\
\left\{ \omega _{tIJ}(\overrightarrow{x}),\mathcal{P}^{tKL}\left( 
\overrightarrow{y}\right) )\right\} &=&\frac{1}{2}(\delta _{I}^{K}\delta
_{J}^{L}-\delta _{I}^{L}\delta _{J}^{K})\delta (\overrightarrow{x}-%
\overrightarrow{y}), \\
\left\{ \omega _{1aIJ}(\overrightarrow{x}),\mathcal{P}_{1}^{bKL}\left( 
\overrightarrow{y}\right) )\right\} &=&P_{1IaJ}^{\text{ \ \ \ \ \ \ \ }%
KbL}\delta (\overrightarrow{x}-\overrightarrow{y})\text{.}
\end{eqnarray*}

The primary constraints are

\[
\pi ^{tN}=0\text{, }\mathcal{P}^{tKL}=0\text{, }\pi ^{bN}=0\text{, }%
C_{1}^{aKL}=\mathcal{P}_{1}^{aKL}-eA^{aKtL}=0 
\]%
and the total fixed Hamiltonian is%
\begin{equation}
\mathcal{H}_{T}^{f}=\int_{\Sigma }(\pi ^{tN}\Lambda _{N}+\mathcal{P}^{tKL}%
\frac{\mathcal{A}_{tKL}}{2}+\pi ^{bN}\Lambda _{bN}+C_{1}^{aKL}\frac{\mathcal{%
A}_{1aKL}}{2})+H_{0}^{f}  \label{RedHamiltonian}
\end{equation}%
where%
\[
H_{0}^{f}\mathcal{=}\int_{\Sigma }(eA^{aKbL}\frac{\Omega (\omega _{1})_{abKL}%
}{2}+eA^{aKtL}D_{1a}^{\omega _{1}}\omega _{tKL}). 
\]

Before we start the analysis of this Hamiltonian, which will be performed
step by step in complete analogy with the treatment of the previous section,
let us precise some remarks concerning the Poisson brackets between the
elements of the fixed phase space. Like for the gauge transformations of $%
\omega _{1aKL}$ (\ref{TransLorentzPROJ1}), to keep the same degrees of
freedom, we project the Poisson brackets acting on the projected elements of
the phase space as 
\[
\left\{ .,\omega _{1aKL}\right\} _{1}=P_{1KaL}^{\text{ \ \ \ \ \ }%
PdQ}\left\{ .,\omega _{1dPQ}\right\} 
\]%
leading to%
\begin{eqnarray}
\left\{ \pi ^{\alpha N},\omega _{1aKL}\right\} _{1} &=&P_{1KaL}^{\text{ \ \
\ \ \ }PbQ}\left\{ \pi ^{\alpha N},\omega _{1dbQ}\right\} =P_{1KaL}^{\text{
\ \ \ \ \ }PdQ}\left\{ \pi ^{\alpha N},P_{1PbQ}^{\text{ \ \ \ \ \ }%
RdS}\omega _{dRS}\right\}  \nonumber \\
&=&P_{1KaL}^{\text{ \ \ \ \ \ }PdQ}\left\{ \pi ^{\alpha N},P_{1PbQ}^{\text{
\ \ \ \ \ }RdS}\right\} \omega _{dRS}  \nonumber \\
&=&P_{1KaL}^{\text{ \ \ \ \ \ }PdQ}\left\{ \pi ^{\alpha N},P_{1PbQ}^{\text{
\ \ \ \ \ }RdS}\right\} \omega _{1dRS}=0  \label{Poisson-Br1-PI-Omrga}
\end{eqnarray}%
where we have used $\omega _{2dRS}=0$ and $P(\delta P)P=0$ true for any
projector $P$. The same computation gives 
\begin{equation}
\left\{ \pi ^{\alpha N}(\overrightarrow{x}),\mathcal{P}_{1}^{aKL}(%
\overrightarrow{y})\right\} _{1}=0  \label{Poisson-Br1-Pi-P}
\end{equation}%
from which we deduce%
\[
\left\{ \pi ^{aN}(\overrightarrow{x}),C_{1}^{bKL}(\overrightarrow{y}%
)\right\} _{1}=-eB^{aNtKbL}\delta (\overrightarrow{x}-\overrightarrow{y})%
\text{.} 
\]

With these projected Poisson brackets the Jacobi identities are satisfied.
For example 
\[
\left\{ \pi ^{\alpha N},\left\{ \omega _{1aIJ},\mathcal{P}_{1}^{bKL}\right\}
_{1}\right\} _{1}+\left\{ \mathcal{P}_{1}^{bKL},\left\{ \pi ^{\alpha
N},\omega _{1aIJ}\right\} _{1}\right\} _{1}+\left\{ \omega _{1aIJ},\left\{ 
\mathcal{P}_{1}^{bKL},\pi ^{\alpha N}\right\} _{1}\right\} _{1}=0 
\]%
as a consequence of (\ref{Poisson-Br1-PI-Omrga}), (\ref{Poisson-Br1-Pi-P})
and 
\begin{eqnarray*}
\left\{ \pi ^{\alpha N},\left\{ \omega _{1aIJ},\mathcal{P}_{1}^{bKL}\right\}
_{1}\right\} _{1} &=&P_{1KaL}^{\text{ \ \ \ \ \ }PdQ}\left\{ \pi ^{\alpha
N},\left\{ \omega _{1dPQ},\mathcal{P}_{1}^{cNM}\right\} \right\} P_{1NcM}^{%
\text{ \ \ \ \ \ }KbL} \\
&=&P_{1KaL}^{\text{ \ \ \ \ \ }PdQ}\left\{ \pi ^{\alpha N},P_{1PdQ}^{\text{
\ \ \ \ \ }NcM}\right\} P_{1NcM}^{\text{ \ \ \ \ \ }KbL}=0\text{.}
\end{eqnarray*}

Now we are ready to perform the treatment of the Hamiltonian (\ref%
{RedHamiltonian}) by using the projected Poisson brackets. The consistency
of the constraint $\pi ^{bN}$ is given by the equation (\ref{EvolC1})
expressed in term of $\omega _{1aKL}$ where $\mathcal{A}_{aKL}$ is replaced
by $\mathcal{A}_{1aKL}$ and whose solution is (\ref{MultAaKL1}) without the
term containing $\Lambda ^{bNc}$. The consistency of the constraint $%
C_{1}^{aKL}$ is given by the same equation (\ref{EvolC2}) where $%
D_{c}eA^{cKaL}$ is replaced by its projection $D_{1c}^{\omega _{1}}eA^{cKaL}$
and whose solution is (\ref{MultiabN}) independently of the condition (\ref%
{TortionConstr}). The substitution of $D_{1c}^{\omega _{1}}eA^{cKaL}$ in (%
\ref{EvolC2}) results from the Poisson bracket of the fixed phase space.

With the new expressions of $\Lambda _{bN}$ and $\mathcal{A}_{1aKL}$ the
Hamiltonian takes the compact form

\[
\mathcal{H}_{T}^{f}=\int_{\Sigma }(\pi ^{tN}\Lambda _{tN}+\mathcal{P}^{tKL}%
\frac{\mathcal{A}_{tKL}}{2}-\mathcal{D}_{t}^{f}-M^{fKL}\frac{\omega _{tKL}}{2%
}) 
\]%
where 
\[
\mathcal{D}_{t}^{f}=-eA^{aKbL}\frac{\Omega (\omega _{1})_{abKL}}{2}-\pi
^{aN}D_{a}^{\omega _{1}}e_{tN}+C_{1}^{aKL}B_{bNtKaL}B^{bNcPdQ}\frac{\Omega
(\omega _{1})_{cdPQ}}{2}\simeq 0 
\]%
and%
\[
\frac{1}{2}M^{fNM}=(D_{a}^{\omega _{1}}\mathcal{(}C_{1}^{aNM}+eA^{aNtM})+%
\frac{1}{2}(\pi ^{\mu N}e_{\mu }^{M}-\pi ^{\mu M}e_{\mu }^{N}))\simeq 0. 
\]

The fixed diffeomorphism constraint is 
\[
\mathcal{D}_{sp}^{f}(\overrightarrow{N})=\dint_{\Sigma }(\pi ^{\mu M}%
\mathcal{L}_{\overrightarrow{N}}(e_{\mu M})+(C_{1}^{aKL}+eA^{aKtL})\mathcal{L%
}_{\overrightarrow{N}}(\omega _{1aKL})). 
\]

A direct computation shows that $\mathcal{D}_{sp}^{f}(\overrightarrow{N})$
satisfies the algebra 
\begin{equation}
\left\{ \mathcal{D}_{sp}^{f}(\overrightarrow{N}),\mathcal{D}_{sp}^{f}(%
\overrightarrow{N^{\prime }})\right\} =\mathcal{D}_{sp}^{f}(\left[ 
\overrightarrow{N},\overrightarrow{N^{\prime }}\right] )\text{.}
\label{fixed-alg1}
\end{equation}

The transformations induced by $\mathcal{D}_{sp}^{f}(\overrightarrow{N})$\
on the primary constraints are given by 
\[
\left\{ \pi ^{\mu N},\mathcal{D}_{sp}^{f}(\overrightarrow{N})\right\} =%
\mathcal{L}_{\overrightarrow{N}}(\pi ^{\mu N})\text{, }\left\{ C_{1}^{aNM},%
\mathcal{D}_{sp}^{f}(\overrightarrow{N})\right\} =\mathcal{L}_{1%
\overrightarrow{N}}(C_{1}^{aNM}) 
\]%
and on the co-tetrad and the spatial components of the dynamic connection by%
\[
\left\{ e_{\mu N},\mathcal{D}_{sp}^{f}(\overrightarrow{N})\right\} =\mathcal{%
L}_{\overrightarrow{N}}(e_{\mu N})\text{, }\left\{ \omega _{1aNM},\mathcal{D}%
_{sp}^{f}(\overrightarrow{N})\right\} =\mathcal{L}_{1\overrightarrow{N}%
}(\omega _{1aNM})\text{.} 
\]

The transformations induced by $\mathcal{M}^{f}(\theta )$ \ on the primary
constraints are given by 
\[
\left\{ \pi ^{\mu N},\mathcal{M}^{f}(\theta )\right\} =\theta _{\text{ \ }%
L}^{N}\pi ^{\mu L}\text{, }\left\{ C_{1}^{aNM},\mathcal{M}^{f}(\theta
)\right\} =\theta _{\text{ \ }L}^{N}C_{1}^{aLM}+\theta _{\text{ \ }%
L}^{M}C_{1}^{aNL} 
\]%
and on the co-tetrad and the spatial components of the dynamic connection by%
\[
\left\{ e_{\mu N},\mathcal{M}^{f}(\theta )\right\} =\theta _{N}^{\text{ \ }%
L}e_{\mu L}\text{, }\left\{ \omega _{1aNM},\mathcal{M}^{f}(\theta )\right\}
=-D_{1a}\theta _{NM} 
\]%
where $\theta _{KL}$ are subject to the condition (\ref{gauge-restrict}).

The above transformations show that the Poisson brackets of $\mathcal{D}%
_{sp}^{f}(\overrightarrow{N})$ and $\mathcal{M}^{f}(\theta )$ with the
primary constraints vanish weakly. Like in the previous section, $\mathcal{D}%
_{t}^{f}$ and $M^{fNM}$ are transformed as scalar densities of weight one by
the diffeomorphisms 
\begin{equation}
\left\{ \mathcal{D}_{sp}^{f}(\overrightarrow{N}),\mathcal{D}%
_{t}^{f}(M)\right\} =\mathcal{D}_{t}^{f}(\mathcal{L}_{\overrightarrow{N}}(M))%
\text{, }\left\{ \mathcal{D}_{sp}^{f}(\overrightarrow{N}),\mathcal{M}(\theta
)\right\} =\mathcal{M}^{f}(\mathcal{L}_{\overrightarrow{N}}(\theta ))\text{.}
\label{fixed-alg3}
\end{equation}%
$\mathcal{D}_{t}^{f}$ is transformed under $\mathcal{M}^{f}(\theta )$ as a
scalare%
\begin{equation}
\left\{ \mathcal{M}^{f}(\theta ),\mathcal{D}_{t}^{f}(M)\right\} =0
\label{fixed-alg5}
\end{equation}%
and $M^{fNM}$ as a tensor%
\[
\left\{ M^{fKL},\mathcal{M}^{f}(\theta )\right\} =\theta _{\text{ }%
N}^{K}M^{fNL}+\theta _{\text{ }N}^{L}M^{fKN} 
\]%
from which we deduce the $so(1,d-1)$ Lie algebra for the generators $M^{fKL}$%
.

The Poisson bracket of $\mathcal{D}_{t}^{f}$ with $\pi ^{aN}(\overrightarrow{%
x})$ gives 
\begin{eqnarray*}
\left\{ \pi ^{aN}(\overrightarrow{x}),\mathcal{D}_{t}^{f}(\overrightarrow{y}%
)\right\} &=& \\
-\frac{1}{2}C_{1}^{eKL}(\overrightarrow{y})\frac{\delta }{\delta e_{aN}(%
\overrightarrow{x})}(B_{bMtKeL}B^{bMcPdQ}(\overrightarrow{y}))\frac{\Omega
_{cdPQ}(\overrightarrow{y})}{2} &\simeq &0
\end{eqnarray*}%
and with $C_{1}^{aNM}(\overrightarrow{x})$ gives 
\begin{eqnarray*}
\left\{ C_{1}^{aNM}(\overrightarrow{x}),\mathcal{D}_{t}^{f}(\overrightarrow{y%
})\right\} &=&D_{1b}^{\omega _{1}}(eA^{bNaM})+eB^{bQtNaM}D_{b}^{\omega
_{1}}e_{tQ} \\
&&+D_{1c}^{\omega _{1}}(C_{1}^{dKL}B_{bNtKdL}B^{bNcNaM}) \\
&=&P_{1KdL}^{\text{ \ \ \ \ \ \ \ }NaM}(eB^{cQbKdL})D_{b}^{\omega _{1}}e_{cQ}
\\
&&+D_{1c}^{\omega _{1}}(C_{1}^{dKL}B_{bNtKdL}B^{bNcNaM}) \\
&=&-B_{ePtKtL}B^{ePtNaM}M^{KL} \\
+D_{1c}^{\omega _{1}}(C_{1}^{dKL}B_{bNtKdL}B^{bNcNaM}) &\simeq &0
\end{eqnarray*}%
which show that the constraint $\mathcal{D}_{t}^{f}$ commutes weakly, in
terms of Poisson brackets, with the primary constraints $\pi ^{aK}$ and $%
C_{1}^{aKL}$. Finally, from a direct calculation we get

\begin{equation}
\left\{ \mathcal{D}_{t}^{f}(M),\mathcal{D}_{t}^{f}(M^{\prime })\right\} =0
\label{fixed-alg4}
\end{equation}%
which shows that the Hamiltonian treatment is coherent and the algebra of
the first-class constraints $\mathcal{M}^{f}(\theta )$, $\mathcal{D}%
_{sp}^{f}(\overrightarrow{N})$ and $\mathcal{D}_{t}^{f}(M)$ closes with
structure constants. The function $M$ and the vector field $\overrightarrow{N%
}$\ may be identified with the usual lapse and shift respectively although
they do not result from the A.D.M. formalism.

In addition of the first-class constraints $\mathcal{M}^{f}(\theta )$, $%
\mathcal{D}_{sp}^{f}(\overrightarrow{N})$ and $\mathcal{D}_{t}^{f}(M)$, we
have the first-class constraints $\pi ^{tN}$ and $\mathcal{P}^{tKL}$ and the
second-class constraints $\pi ^{aN}$ and $C_{1}^{aNM}$. The physical degrees
of freedom per point in space-time are obtained by subtracting from the $%
d(5d-3)$\ degrees of freedom of the fixed phase space the number $2d(d-1)$
of the second-class constraints and twice the number $d(d+1)$ of the
first-class constraints to get $d(d-3)$ which is exactly the number of the
degrees of freedom of the physical phase space of the d-dimensional general
relativity.

Remark: in order to avoid the constraint (\ref{TortionConstr}), one might
wonder what happen if instead of fixing the non dynamic part of the
connection to zero one solve its equation of motion and then put the
solution back into the action as in \cite{Contreras}. Since the
decomposition of the connection in dynamic and non-dynamic part is unique,
the part of solution of the zero-torsion \cite{Peldan} which corresponds to $%
\omega _{2aKL}$ must be of the form

\[
\omega _{2aKL}=P_{2aKL}^{\text{ \ \ \ \ \ \ }bPQ}e_{\mu P}\nabla
_{b}e_{Q}^{\mu }=P_{2aKL}^{\text{ \ \ \ \ \ \ }bPQ}e_{\mu P}(\partial
_{b}e_{Q}^{\mu }+\Gamma _{b\nu }^{\mu }e_{Q}^{\nu }) 
\]%
where $\nabla _{a}e_{M}^{\mu }=\partial _{a}e_{M}^{\mu }+\Gamma _{a\nu
}^{\mu }e_{M}^{\nu }$ is the covariant derivative with respect to
Christoffel's symbol $\Gamma _{a\nu }^{\mu }$ whose expression contains
linearly the derivatives of the components of the co-tetrad. an explicit
computation shows that $\omega _{2aKL}$ contains linearly only the time
derivatives of the spatial components of the co-tetrad. This leaves $\pi
^{tN}$ as a primary constraint but not $\pi ^{aN}$ which takes the form

\[
\pi ^{aN}=\left( D_{b}^{\omega _{1}}eA^{bKaL}+\omega _{2b\text{ \ \ }M}^{%
\text{ \ \ \ }K}eA^{bMaL}+\omega _{2b\text{ \ \ }M}^{\text{ \ \ \ }%
L}eA^{bKaM}\right) \frac{\delta \omega _{2aKL}}{\delta \partial _{t}e_{aN}} 
\]%
where $\frac{\delta \omega _{2aKL}}{\delta \partial _{t}e_{aN}}=(P_{2aKL}^{%
\text{ \ \ \ \ \ \ }bPQ}e_{\mu P}\frac{\delta }{\delta \partial _{t}e_{aN}}%
(\Gamma _{a\nu }^{\mu })e_{Q}^{\nu }$ is not null. The linear dependence of $%
\pi ^{aN}$ with respect to the time derivative of the spatial components of
the co-tetrad leads to a second order formalism.

\section{\protect\bigskip The algebra of constraints in terms of Dirac
brackets}

\bigskip

In the previous section we have showed that the set of constraints is
complete and closed meaning that the total Hamiltonian $\mathcal{H}_{T}^{f}$
(\ref{RedHamiltonian}) is consistent. In this section we consider the
second-class constraints $\pi ^{aN}$ and $C_{1}^{aKL}$ as strong equalities
by eliminating them. In this case the algebra of the first-class constraints
must be computed with the projected Dirac brackets defined in terms of the
Poisson brackets of the previous section as

\[
\left\{ A,B\right\} _{D}=\left\{ A,B\right\} _{1}-\left\{ A,C_{i}\right\}
_{1}\left\{ C_{i},C_{j}\right\} _{1}^{-1}\left\{ C_{j},B\right\} _{1} 
\]%
where $C_{i}=\left( \pi ^{aN},C_{1}^{aKL}\right) $. The inverse of the
Poisson bracket 
\[
\left\{ \pi ^{bN}(\overrightarrow{x}),C_{1}^{aKL}(\overrightarrow{y}%
)\right\} _{1}\simeq eB^{bNaKtL}\delta (\overrightarrow{x}-\overrightarrow{y}%
)=-eB^{bNtKaL}\delta (\overrightarrow{x}-\overrightarrow{y}) 
\]%
is given by $\left\{ \pi ^{bN},C_{1}^{aKL}\right\}
_{1}^{-1}=e^{-1}B_{bNtKaL} $ satisfying%
\[
\left\{ \pi ^{bN},C_{1}^{aKL}\right\} _{1}^{-1}\left\{ C_{1}^{aKL},\pi
^{cM}\right\} _{1}=\delta _{b}^{c}\delta _{N}^{M} 
\]%
and%
\[
\left\{ C_{1}^{aKL},\pi ^{bN}\right\} _{1}^{-1}\left\{ \pi
^{bN},C_{1}^{dPQ}\right\} _{1}=P_{1KaL}^{\text{ \ \ \ \ \ }PdQ}\text{.} 
\]

Now we can consider the constraints of second-class $\pi ^{bN}$ and $%
C_{1}^{aKL}$ as strong equalities by eliminating them from the total
Hamiltonian $\mathcal{H}_{T}^{\prime }$ to get the reduced Hamiltonian

\begin{equation}
\mathcal{H}_{T}^{r}=\int_{\Sigma }(\pi ^{tN}\Lambda _{tN}+\mathcal{P}^{tKL}%
\frac{\mathcal{A}_{tKL}}{2}-\mathcal{D}_{t}^{r}-M^{rKL}\frac{\omega _{tKL}}{2%
})  \label{HamiltonianFinal}
\end{equation}%
where 
\[
M^{rKL}=(\pi ^{tK}e_{t}^{L}-\pi ^{tL}e_{t}^{K})+2D_{a}^{\omega
_{1}}(eA^{aKtL}) 
\]%
and%
\begin{equation}
\mathcal{D}_{t}^{r}=-eA^{aKbL}\frac{\Omega (\omega _{1})_{abKL}}{2}\text{.}
\label{DiffeoFinal}
\end{equation}

The diffeomorphism constraint reduces to%
\[
\mathcal{D}_{sp}^{r}(\overrightarrow{N})=\dint_{\Sigma }\left( \pi ^{tN}%
\mathcal{L}_{\overrightarrow{N}}(e_{tN})+eA^{aKtL}\mathcal{L}_{%
\overrightarrow{N}}(\omega _{1aKL})\right) \text{.} 
\]

The Hamiltonian (\ref{HamiltonianFinal}) is defined on the reduced phase
space $e_{aN}$, $\omega _{1aKL}$, $e_{tN\text{ }}$, $\pi ^{tK}$, $\omega
_{tKL}$, and $\mathcal{P}^{tKL}$ equipped with the following non zero Dirac
brackets:

\begin{eqnarray*}
\left\{ e_{tN}(\overrightarrow{x}),\pi ^{tM}(\overrightarrow{y})\right\}
_{D} &=&\delta _{N}^{M}\delta (\overrightarrow{x}-\overrightarrow{y})\text{, 
} \\
\left\{ \omega _{tIJ}(\overrightarrow{x}),\mathcal{P}^{tKL}\left( 
\overrightarrow{y}\right) )\right\} _{D} &=&\frac{1}{2}(\delta
_{I}^{K}\delta _{J}^{L}-\delta _{I}^{L}\delta _{J}^{K})\delta (%
\overrightarrow{x}-\overrightarrow{y})
\end{eqnarray*}%
and

\begin{equation}
\left\{ e_{aN}(\overrightarrow{x}),\omega _{1bKL}(\overrightarrow{y}%
)\right\} _{D}=e^{-1}B_{aNtKbL}\delta (\overrightarrow{x}-\overrightarrow{y})%
\text{.}  \label{Dirac-Brac-E-Ome}
\end{equation}

Note that as opposite to the results obtained in \cite{Alexandrov}, the
dynamic connection is Dirac self commuting as a consequence of (\ref%
{Poisson-Br1-PI-Omrga}).

These projected Dirac brackets guarantee the Jacobi identities. In fact for
the non trivial example%
\[
\left\{ e_{cN},\left\{ \omega _{1aKL},\omega _{1bPQ}\right\} _{D}\right\}
_{D}+\left\{ \omega _{1bPQ},\left\{ e_{cN},\omega _{1aKL}\right\}
_{D}\right\} _{D}+\left\{ \omega _{1aKL},\left\{ \omega
_{1bPQ},e_{cN}\right\} _{D}\right\} _{D} 
\]%
the first term vanishes and the second and third term give

\begin{eqnarray*}
&-&P_{1Kal}^{\text{ \ \ \ \ \ }RdS}\frac{\delta }{\delta e_{hI}}%
(e^{-1}B_{cNtRdS})e^{-1}B_{hItPbQ}+P_{1PbQ}^{\text{ \ \ \ \ \ }RdS}\frac{%
\delta }{\delta e_{hI}}(e^{-1}B_{cNtRdS})e^{-1}B_{hItKaL} \\
&=&e^{hI}e^{-1}B_{cNtKaL}e^{-1}B_{hItPbQ}-e^{hI}e^{-1}B_{cNtPbQ}e^{-1}B_{hItKaL}
\\
&&-P_{1Kal}^{\text{ \ \ \ \ \ }RdS}e^{-1}\frac{\delta }{\delta e_{hI}}%
(B_{cNtRdS})e^{-1}B_{hItPbQ}+e^{-1}P_{1PbQ}^{\text{ \ \ \ \ \ }RdS}\frac{%
\delta }{\delta e_{hI}}(B_{cNtRdS})e^{-1}B_{hItKaL}
\end{eqnarray*}%
where we have used 
\begin{eqnarray*}
\left\{ \omega _{1aKL},e^{-1}B_{cNtPbQ}\right\} _{D} &=&P_{1PbQ}^{\text{ \ \
\ \ \ }RdS}\frac{\delta }{\delta e_{hI}}(e^{-1}B_{cNtRdS})\left\{ \omega
_{1aKL},e_{hI}\right\} _{D} \\
&=&-P_{1PbQ}^{\text{ \ \ \ \ \ }RdS}\frac{\delta }{\delta e_{hI}}%
(e^{-1}B_{cNtRdS})e^{-1}B_{hItKaL}
\end{eqnarray*}%
and $\frac{\delta }{\delta e_{hI}}e^{-1}=-e^{-1}e^{hI}$. A direct
computation leads to%
\begin{eqnarray*}
&&+e^{-1}P_{1PbQ}^{\text{ \ \ \ \ \ }RdS}\frac{\delta }{\delta e_{hI}}%
(B_{cNtRdS})e^{-1}B_{hItKaL}-P_{1Kal}^{\text{ \ \ \ \ \ }RdS}e^{-1}\frac{%
\delta }{\delta e_{hI}}(B_{cNtRdS})e^{-1}B_{hItPbQ} \\
&=&-e^{hI}e^{-1}B_{cNtKaL}e^{-1}B_{hItPbQ}+e^{hI}e^{-1}B_{cNtPbQ}e^{-1}B_{hItKaL}
\end{eqnarray*}%
showing that the Jacobi identities are satisfied.

Now we are ready to calculate the algebra of constraints in terms of the
projected Dirac brackets. The smeared Lorentz Constraint%
\[
\mathcal{M}^{r}(\theta )=\int_{\Sigma }((\pi _{1}^{tK}e_{t}^{L}-\pi
_{1}^{tL}e_{t}^{K})+2D_{a}^{\omega _{1}}eA^{aKtL})\frac{\theta _{KL}}{2}%
=\int_{\Sigma }M^{rKL}\frac{\theta _{KL}}{2} 
\]%
acts on $e_{\mu N}$ and $\omega _{1aKL}$ like local infinitesimal
transformations of gauge 
\[
\left\{ e_{\mu N},\mathcal{M}^{r}(\theta )\right\} _{D}=\theta _{N}^{\text{
\ }M}e_{\mu M}\text{, \ }\left\{ \omega _{1aKL},\mathcal{M}^{r}(\theta
)\right\} _{D}=-D_{1a}\theta _{KL} 
\]%
leading to%
\[
\left\{ \mathcal{M}^{r}(\theta ),\mathcal{D}_{t}^{r}(M)\right\} _{D}=0 
\]%
as a consequence of (\ref{Inv-H0fixed}) and to%
\[
\left\{ M^{rKL},\mathcal{M}^{r}(\theta )\right\} _{D}=\theta _{\text{ }%
N}^{K}M^{rNL}+\theta _{\text{ }N}^{L}M^{rKN} 
\]%
from which we get the $so(1,d-1)$ Lie algebra for the generators $M^{rKL}$.

From a direct computation we get for the spatial diffeomorphism constraint%
\[
\left\{ \mathcal{D}_{sp}^{r}(\overrightarrow{N}),\mathcal{D}_{sp}^{r}(%
\overrightarrow{N^{\prime }})\right\} _{D}=\mathcal{D}_{sp}^{r}(\left[ 
\overrightarrow{N},\overrightarrow{N^{\prime }}\right] ). 
\]

The transformations induced by $\mathcal{D}_{sp}^{r}(\overrightarrow{N})$ on 
$e_{\mu N}$ and $\omega _{1aKL}$ are given by%
\[
\left\{ e_{\mu N},\mathcal{D}_{sp}^{r}(\overrightarrow{N})\right\} _{D}=%
\mathcal{L}_{\overrightarrow{N}}(e_{\mu N})\text{, \ }\left\{ \omega _{1aKL},%
\mathcal{D}_{sp}^{r}(\overrightarrow{N})\right\} _{D}=\mathcal{L}_{1%
\overrightarrow{N}}(\omega _{1aKL}). 
\]

In view of these transformations we then deduce that $\mathcal{D}_{t}^{r}$
and $M^{rKL}$ transform under the spatial diffeomorphisms as scalar
densities of weight one leading to%
\[
\left\{ \mathcal{D}_{sp}^{r}(\overrightarrow{N}),\mathcal{D}%
_{t}^{r}(M)\right\} _{D}=\mathcal{D}_{t}^{r}(\mathcal{L}_{\overrightarrow{N}%
}(M))\text{, }\left\{ \mathcal{D}_{sp}^{r}(\overrightarrow{N}),\mathcal{M}%
^{r}(\theta )\right\} _{D}=\mathcal{M}^{r}(\mathcal{L}_{\overrightarrow{N}%
}(\theta ))\text{.} 
\]

Finally, a direct calculation gives for the smeared scalar constraint 
\begin{eqnarray*}
\left\{ \mathcal{D}_{t}^{r}(M),\mathcal{D}_{t}^{r}(M^{\prime })\right\} _{D}
&=&\left\{ \int_{\Sigma }MeA^{aKbL}\frac{\Omega (\omega _{1})_{abKL}}{2}%
,\int_{\Sigma }M^{\prime }eA^{cNdM}\frac{\Omega (\omega _{1})_{cdNM}}{2}%
\right\} _{D} \\
&=&\int_{\Sigma }D_{a}(MeA^{aKbL})M^{\prime }eB^{hQcNdM}e^{-1}B_{hQtKbL}%
\frac{\Omega (\omega _{1})_{cdNM}}{2} \\
&&-\int_{\Sigma }D_{c}(M^{\prime }eA^{cNdM})MeB^{hQaKbL}e^{-1}B_{hQtNdM}%
\frac{\Omega (\omega _{1})_{abKL}}{2} \\
&=&\int_{\Sigma }MM^{\prime }D_{a}(eA^{aKbL})eB^{hQcNdM}e^{-1}B_{hQtKbL}%
\frac{\Omega (\omega _{1})_{cdNM}}{2} \\
&&-\int_{\Sigma }MM^{\prime }D_{c}(eA^{cNdM})eB^{hQaKbL}e^{-1}B_{hQtNdM}%
\frac{\Omega (\omega _{1})_{abKL}}{2} \\
&=&0
\end{eqnarray*}%
where we have used 
\begin{eqnarray*}
&&\int_{\Sigma }\partial _{a}(M)eA^{aKbL}B^{hQcNdM}B_{hQtKbL}\frac{\Omega
(\omega _{1})_{cdNM}}{2} \\
&=&\int_{\Sigma }\partial _{a}(M)eA^{aKbL}\frac{1}{2(d-2)}%
e_{tQ}A_{bKhL}B^{hQcNdM}\frac{\Omega (\omega _{1})_{cdNM}}{2}=0
\end{eqnarray*}%
due to $A^{aKbL}A_{tKcL}=0$ and $e_{tQ}B^{hQcNdM}=0$. The above Dirac
brackets between $\mathcal{D}_{sp}^{r}(\overrightarrow{N})$, $\mathcal{M}%
^{r}(\theta )$ and $\mathcal{D}_{t}^{r}(M)$ show that the algebra of the
reduced first-class constraints closes with structure constants.

As in the previous section, the degrees of freedom of the physical phase
space is $d(d-3)$ which is exactly the number of the degrees of freedom of
the physical phase space of the $d-$dimensional general relativity.

In terms of Dirac brackets, the Hamiltonian (\ref{HamiltonianFinal})
propagates the phase space variable from an initial hypersurface $\Sigma
_{t_{0}}$ to the hypersurface $\Sigma _{t_{0}+\delta t}$. In this sense, we
can interpret geometrically that the Hamiltonian (\ref{HamiltonianFinal})
propagates the hypersurface $\Sigma _{t_{0}}$ in the space-time. This
propagation is done along the infinitesimal time $\delta t$ or $M\delta t$
depending on whether we consider the scalar density $\mathcal{D}_{t}^{r}$ or 
$\mathcal{D}_{t}^{r}(M)$ in the Hamilton equation while the Lorentz
constraint induces, during this propagation, a gauge transformation of
infinitesimal parameter $\theta _{KL}=\delta t\omega _{tKL}$. The constraint
(\ref{DiffeoFinal}) acts independently of (\ref{HamiltonianFinal}) and
induces diffeomorphisms on each hypersurfaces.

We end this section by solving Hamilton's equations in terms of Dirac
brackets. For the co-tetrad components $e_{cN}$, we get

\begin{eqnarray*}
\partial _{t}e_{cN}(x) &=&\left\{ e_{cN}(x),\mathcal{H}_{T}^{r}\right\}
_{D}=-D_{a}^{\omega _{1}}(eA^{aKbL})e^{-1}B_{cNtKbL} \\
&&+eA^{aKtL}(e^{-1}B_{cNtKaM}\omega _{t\text{ \ }L}^{M}+e^{-1}B_{cNtLaM}%
\omega _{tK}^{\text{ \ \ }M})\text{.}
\end{eqnarray*}

The first term of the right hand side gives

\begin{eqnarray*}
-D_{a}^{\omega _{1}}(eA^{aKbL})e^{-1}B_{cNtKbL} &=&-e^{-1}B_{cNtKbL}eB^{\mu
MaKbL}D_{a}^{\omega _{1}}e_{\mu M} \\
&=&D_{c}^{\omega _{1}}e_{tM}-eB_{cNtKbL}e^{-1}B^{dMaKbL}D_{a}^{\omega
_{1}}e_{dM} \\
&=&D_{c}^{\omega _{1}}e_{tM}+e^{-1}B_{cNtKtL}M^{KL}
\end{eqnarray*}%
and a direct computation gives for the second term of the right hand side%
\[
eA^{aKtL}(e^{-1}B_{cNtKaM}\omega _{t\text{ \ }L}^{M}+e^{-1}B_{cNtLaM}\omega
_{tK}^{\text{ \ \ }M})=-\omega _{tN}^{\text{ \ \ }M}e_{cM} 
\]%
leading to 
\[
D_{t}e_{aN}-D_{a}^{\omega _{1}}e_{tN}=B_{cNtKtL}M^{KL}\text{.} 
\]

Since $\omega _{tKL\text{ }}$can be considered as Lagrange multiplier, in
addition to the above equation we have%
\[
B^{dMaKtL}D_{a}e_{dM}=D_{a}^{\omega _{1}}\left( eA^{aKtL}\right) =M^{KL}=0 
\]%
leading 
\[
D_{t}e_{aN}-D_{a}^{\omega _{1}}e_{tN}=0. 
\]

The solutions of $D_{a}^{\omega _{1}}\left( eA^{aKtL}\right) =M^{KL}=0$ are $%
\omega _{aKL}^{s}=e_{\mu N}\nabla _{a}e_{M}^{\mu }=e_{\mu N}(\partial
_{a}e_{M}^{\mu }+\Gamma _{a\nu }^{\mu }e_{M}^{\nu })$. The solution $\omega
_{aKL}^{s}$ is injected in the second equation to get%
\begin{eqnarray*}
D_{t}e_{aN}-D_{a}^{\omega ^{s}}e_{tN} &=&\partial _{t}e_{aN}+\omega
_{tNM}e_{a}^{M}-\partial _{a}e_{tN}-\omega _{aNM}^{s}e_{t}^{M} \\
&=&\partial _{t}e_{aN}+\omega _{tNM}e_{a}^{M}-\partial _{a}e_{tN}+\nabla
_{a}e_{\mu N}e_{M}^{\mu }e_{t}^{M} \\
&=&\partial _{t}e_{aN}+\omega _{tNM}e_{a}^{M}-\Gamma _{at}^{\mu }e_{\mu M}
\end{eqnarray*}%
leading to $\omega _{tNM}=e_{\mu N}\nabla _{t}e_{M}^{\mu }=e_{\mu
N}(\partial _{t}e_{M}^{\mu }+\Gamma _{t\nu }^{\mu }e_{M}^{\nu })$ \cite%
{Peldan} which exhibits solutions of the zero torsion, condition to have an
equivalence between the tetrad-connection gravity and the general relativity.

\section{Conclusion}

\bigskip

The results presented in this paper show that a modified action of the
tetrad-gravity where the non-dynamic connection $\omega _{2aKL}$ is fixed to
zero makes possible the construction of a consistent Hamiltonian formulation
for any dimension $d\geq 3$ without Barbero-Immirzi's parameter neither the
A.D.M.\ decomposition of the action.

Unlike the works where the ADM decomposition of action is taken as starting
point leading to a Hamiltonian system where the algebra of the first-class
constraints closes with structure functions we have showed that, by starting
from a phase space without using the A.D.M. decomposition, we get a coherent
Hamiltonian formalism with an algebra of the first-class constraints which
closes with structure constants. This algebra expresses the invariance of
the action under a true Lie group whose generators are the first class
constraints.

The absence of structure functions is due to the fact that the scalar
function $M$ and the spatial vector field $\overrightarrow{N}$ are
introduced as test functions independently of the co-tetrad not as objects
which result from the A.D.M. decomposition of the tangent space where the
pure deformation of the space-like hypersurface, expressed in term of the
lapse, requires for its definition the metric of space-time which appears in
the structure functions \cite{Hojman} . It was shown in \cite{Bergmann} that
it is possible to obtain an algebra of the diffeomorphism constraints which
closes with structure constants by considering the general transformations
of the coordinates which depend on the metric.

In \cite{Bodendorfer} the simplicity constraint, which corresponds to the
primary constraint $C^{aKL}$ of the section 2, is split into boost and
non-boost part while in our case the decomposed is done by the projectors
like $C^{aKL}=C_{1}^{aKL}+C_{2}^{aKL}$ where $C_{1}^{aKL}$ is the primary
constraint of the section 3 and $C_{2}^{aKL}=\mathcal{P}_{2}^{aKL}=\mathcal{P%
}^{cNM}P_{2NcM}^{\text{ \ \ \ \ \ }KaL}$ is the conjugate momenta of the
non-dynamic part of the connection which is a primary constraint whose the
consistency condition is (\ref{TortionConstr}). The fixing in the action of
the non-dynamic part of connection to zero allowed us to eliminate the
constraint (\ref{TortionConstr}) to get a coherent Hamiltonian formalism.

Note that all reduced first-class constraints are polynomial but the phase
space variables obey a non polynomial Dirac bracket because of the presence
of $e^{-1}$ (\ref{Dirac-Brac-E-Ome}). Since the rank of the projector $%
P_{1NcM}^{\text{ \ \ \ \ \ }KaL}$ is $d(d-1)$, the number of the
independents components of the dynamic connection is equal to the ones of
the co-tetrad $e_{aK}.$ This allows us to perform an invertible
transformation%
\[
\omega _{1aKL}\longrightarrow \mathcal{P}^{bN}=eB^{bNtKaL}\omega
_{1aKL}\Longleftrightarrow \omega _{1aKL}=\mathcal{P}^{bN}e^{-1}B_{bNtKaL} 
\]%
to get a reduced phase space obeying the following canonical commutation
relations in terms of Dirac brackets%
\begin{eqnarray}
\left\{ e_{aN}(\overrightarrow{x}),\mathcal{P}^{bM}(\overrightarrow{y}%
)\right\} _{D} &=&\delta _{a}^{b}\delta _{N}^{M}\delta (\overrightarrow{x}-%
\overrightarrow{y}),  \nonumber \\
\left\{ e_{aN}(\overrightarrow{x}),e_{bM}(\overrightarrow{y})\right\} _{D}
&=&0\text{ and }\left\{ \mathcal{P}^{aN}(\overrightarrow{x}),\mathcal{P}%
^{bM}(\overrightarrow{y})\right\} _{D}=0\text{.}
\label{caninical-Dirac-Ph-sp}
\end{eqnarray}

Substituting $\omega _{1aKL}$ by $\mathcal{P}^{bN}e^{-1}B_{bNtKaL}$, we get
polynomial constraints for 
\[
M^{rKL}=\partial _{a}eA^{aKtL}+\frac{1}{2}(\mathcal{P}^{aK}e_{a}^{L}-%
\mathcal{P}^{aL}e_{a}^{K}) 
\]%
and%
\[
\mathcal{D}_{sp}^{r}(\overrightarrow{N})=\dint_{\Sigma }\left( \pi ^{tN}%
\mathcal{L}_{\overrightarrow{N}}(e_{tN})+\mathcal{P}^{aN}\mathcal{L}_{%
\overrightarrow{N}}(e_{aN})\right) 
\]%
whereas the scalar constraint becomes non-polynomial because of the
quadratic term of the dynamic connection $\omega _{1aKL}(e,\mathcal{P})=%
\mathcal{P}^{bN}e^{-1}B_{bNtKaL}$ in the curvature $\Omega (\omega
_{1})_{abKL}$. From the same calculation done in section 4 to check the
Jacobi identities we deduce%
\[
\left\{ \omega _{1aKL},\omega _{1dPQ}\right\} _{D}=\left\{ \mathcal{P}%
^{bN}e^{-1}B_{bNtKaL},\mathcal{P}^{cM}e^{-1}B_{cMtPdQ}\right\} _{D}=0 
\]%
conducting, with the use of (\ref{caninical-Dirac-Ph-sp}), to the same
algebra of first class constraints of the previous section. A similar
approach with the same phase space was adopted in \cite{Contreras} where the
connection is decomposed in dynamic and non-dynamic part and where the
latter is replaced by the solution of its equation of motion. If the algebra
of the lorentz constraints is explicitly computed, because they depend only
on the dynamic part of the connection, the computation of the bracket
between the translation constraints raises problems because of the reasons
given in the remark of the section 3.

In three dimension $eB^{bNaKcL}$ vanishes, as a result of the antisymmetry
of the spacial indices $b,$ $a$ and $c$, implying that the projected
connection $\omega _{2aKL}$ disappears in the same way. All the previous
results are valid in the three dimensional case . In addition, the first
class constraint $P^{N}$ can be retained\ instead of $\mathcal{D}_{t}(M)$
and $\mathcal{D}_{sp}(\overrightarrow{N})$. $P^{N}$ and $M^{KL}$ \ obey the
following Dirac bracket

\begin{eqnarray}
\left\{ M^{NM}(\overrightarrow{x}),M^{KL}(\overrightarrow{y})\right\} _{D}
&=&(\eta ^{NL}M^{MK}(\overrightarrow{x})+\eta ^{MK}M^{NL}(\overrightarrow{x})
\nonumber \\
&&-\eta ^{NK}M^{ML}(\overrightarrow{x})-\eta ^{ML}M^{NK}(\overrightarrow{x}%
))\delta (\overrightarrow{x}-\overrightarrow{y}),  \label{Poincare-Lorentz1}
\end{eqnarray}

\begin{equation}
\left\{ P^{N}(\overrightarrow{x}),M^{KL}(\overrightarrow{y})\right\}
_{D}=\left( \eta ^{NL}P^{K}(\overrightarrow{x})-\eta ^{NK}P^{L}(%
\overrightarrow{y})\right) \delta (\overrightarrow{x}-\overrightarrow{y})
\label{Poicare-Lorentz2}
\end{equation}%
and

\begin{equation}
\left\{ P^{N}(\overrightarrow{x}),P^{M}(\overrightarrow{y})\right\} _{D}=0
\label{Poincare-Lorentz3}
\end{equation}%
which exhibit the Lie algebra of the Poincar\'{e} group, where $\delta
te_{tN}$ plays the role of the infinitesimal local translation. This shows
that, in the three dimensional case, we can either consider the Poincar\'{e}
group or the Lorentz and diffeomorphism group. The physical degrees of
freedom vanish, expressing the topological character of the $d=3$ gravity.

We end this paper by noticing that, as opposed to \cite{Frolov} where it is
claimed that diffeomorphism invariance is not a gauge symmetry derived from
the first-class constraints of the tetrad-gravity or in \cite{Witten} where
the equivalence between the translation transformations of Poincar\'{e}
group and the diffeomorphism can only be established on-shell. In this
paper, we have shown that the symmetries of the Hamiltonian formalism of
three dimensional tetrad-connection gravity are obtained either through the
first-class constraints of Poincar\'{e} group or the ones of Lorentz and
diffeomorphisms which show that this equivalence is established off-shell.

For $d\geq 4$, the smeared first-class constraint $P^{N}$can be written as a
sum of first-class constraints%
\[
P(N)=\dint N_{N}eB^{tNaKbL}\frac{\Omega _{abKL}}{2}=-\mathcal{D}_{t}(N)+%
\mathcal{D}_{sp}(\overrightarrow{N})+\mathcal{M}^{r}(\theta ) 
\]%
where $\mathcal{D}_{sp}(\overrightarrow{N})=\dint_{\Sigma }eA^{aKtL}\mathcal{%
L}_{\overrightarrow{N}}(\omega _{1aKL})$, $\mathcal{M}^{r}(\theta
)=\int_{\Sigma }D_{a}^{\omega _{1}}eA^{aKtL}\theta _{KL}$, $N_{N}$ is a
Lorentzian vector, $N=N_{N}e^{tN}$, $\theta _{KL}=N_{N}e^{aN}\omega _{aKL}$
and $N_{N}e^{aN}$ are the $d-1$ components of the vector $\overrightarrow{N}$
tangent to $\Sigma _{t}$. The Dirac bracket between the smeared translation
constraints does not vanish but gives a sum of which a part is linear in the
first-class constraints with structure functions and the other quadratic.
The quadratic part results from the fact that the functions $N$, $%
\overrightarrow{N}$ and $\theta _{KL}$ depend on the components of the
tetrad and thus contribute to the results of Dirac's bracket. This shows
that $P^{N}$ does not satisfy (\ref{Poincare-Lorentz3}) and therefore does
not correspond to the translational part of the Lie algebra of the Poincar%
\'{e} group. Only the Lorentz group and the diffeomorphisms are symmetries,
not the translation part of the Poincar\'{e} group, contrary to what is
claimed in \cite{Kuzmin} .

\section{Acknowledgements}

M. Lagraa would like to thank Michel Dubois-Violette for helpful discussions.

\end{document}